\documentclass[nonacm,acmsmall]{acmart}

\setcopyright{acmlicensed}
\acmDOI{10.1145/3704901}
\acmYear{2025}
\acmJournal{PACMPL}
\acmVolume{9}
\acmNumber{POPL}
\acmMonth{1}

\usepackage{xcolor,colortbl}
\usepackage{amsmath}
\usepackage{graphicx}
\usepackage{booktabs}
\usepackage{latexsym}	
\usepackage{array}		
\usepackage{multirow}
\usepackage{subcaption}
\usepackage{algorithm}
\usepackage{algpseudocode}
\usepackage{paralist}   
\usepackage{xspace}
\usepackage{adjustbox}
\usepackage{balance}
\usepackage{wasysym}
\usepackage{rotating}   
\usepackage{listings}
\usepackage{pifont}
\usepackage{framed}
\usepackage[shortlabels]{enumitem}
\usepackage{microtype}
\usepackage{marvosym}
\usepackage{amsmath, amsthm}

\newcommand{\myparagraph}[1]{\vspace{1ex}\noindent{\textbf{#1.}}}

\newcommand{\revised}[1]{{\color{black}{#1}}}

\definecolor{mygreen}{rgb}{0,0.6,0}
\definecolor{mygray}{rgb}{0.5,0.5,0.5}
\definecolor{mymauve}{rgb}{0.58,0,0.82}
\definecolor{darkblue}{rgb}{0.0,0.0,0.6}
\definecolor{maroon}{RGB}{102, 0, 0}
\definecolor{Maroon}{cmyk}{0,0.87,0.68,0.32}
\definecolor{darkred}{RGB}{139, 0, 0}
\definecolor{forestgreen}{RGB}{34, 139, 34}

\begin{document}

\title{An Incremental Algorithm for Algebraic Program Analysis}

\author{Chenyu Zhou}
\orcid{0009-0006-8493-6886}
\affiliation{%
  \institution{University of Southern California}
  \city{Los Angeles}
  \country{USA}
}
\email{czhou691@usc.edu}

\author{Yuzhou Fang}
\orcid{0000-0002-4933-7443}
\affiliation{%
  \institution{University of Southern California}
  \city{Los Angeles}
  \country{USA}
}
\email{yuzhou.fang@usc.edu}

\author{Jingbo Wang}
\orcid{0000-0001-5877-2677}
\affiliation{%
  \institution{Purdue University}
  \city{West Lafayette}
  \country{USA}
}
\email{wang6203@purdue.edu}

\author{Chao Wang}
\orcid{0009-0003-4684-3943}
\affiliation{%
  \institution{University of Southern California}
  \city{Los Angeles}
  \country{USA}
}
\email{wang626@usc.edu}



\begin{abstract}
We propose a  method  for conducting algebraic program analysis (APA) incrementally in response to changes of the program under analysis. 
APA is a program analysis paradigm that consists of two distinct steps: computing a path expression that succinctly summarizes the set of program paths of interest, and interpreting the path expression using a properly-defined semantic algebra to obtain program properties of interest. 
In this context, the goal of an incremental algorithm is to reduce the analysis time by leveraging the intermediate results computed before the program changes. 
We have made two main contributions.  
First, we propose a data structure for efficiently representing path expression as a tree together with a tree-based interpreting method.
Second, we propose techniques for efficiently updating the program properties in response to changes of the path expression. 
We have implemented our method and evaluated it on thirteen Java applications from the DaCapo benchmark suite. 
The experimental results show that both our method for incrementally computing path expression and our method for incrementally interpreting path expression are effective in speeding up the analysis. Compared to the baseline APA and two state-of-the-art APA methods,  the speedup of our method ranges from 160$\times$ to 4761$\times$ depending on the types of program analyses performed. 
\end{abstract}

\maketitle

\section{Introduction}

Algebraic program analysis (APA) is a general framework for analyzing the properties  of a computer program at various levels of abstraction.  \revised{At a high level, it can be viewed as an alternative to the classic, chaotic-iteration based program analysis.} 
While both iterative program analysis and APA view the space of program properties (or facts) of interest as an abstract structure, i.e., a lattice or a semi-lattice, the way they compute these properties are different.
Iterative program analysis follows an \emph{interpret-and-then-compute} approach, meaning that it first interprets the semantics of a program using a properly-defined abstract domain and an abstract transformer, and then computes the properties by propagating them through the control flow graph iteratively, until a fixed point is reached.  Examples include the unified data-flow analysis framework of Kildall~\cite{Kildall73} and abstract interpretation of Cousot and Cousot~\cite{CousotC77}.
In contrast, APA follows a \emph{compute-and-then-interpret} approach, meaning that it first computes a so-called \emph{path expression}, which is a type of regular expression for summarizing all program paths of interest, and then interprets the path expression using a properly-defined semantic algebra and a semantic function to compute program properties of interest.

The \emph{compute-and-then-interpret} approach of APA exploits the fact that the semantics of a program is determined by its structure and the semantics of its components.  In other words, the approach is inherently compositional.  While perhaps not being as widely-used as iterative program analysis, APA has a long history that can be traced back to an algebraic approach for solving path problems in directed graphs, for which Tarjan~\cite{tarjan1981fast} proposed a fast algorithm and a unified framework~\cite{tarjan1981unified}; in addition to solving graph-theoretic problems such as computing single-source shortest paths, the framework is able to solve many program analysis problems. 
%
\revised{
Being compositional allows APA to scale to large programs,  to be applied to incomplete programs, and to be parallelized easily~\cite{kincaid2021algebraic}. 
}%
Due to these reasons, APA has been employed in many settings.  For example, beyond classic data-flow analyses,  APA has been used for invariant generation~\cite{KincaidCBR18}, termination analysis~\cite{ZhuKFY21}, predicate abstraction~\cite{RepsTP16}, and more recently, for analyzing probabilistic programs~\cite{WangHR18}.

However, we are not aware of any existing algorithm for conducting APA incrementally in response to small and frequent changes of the program. 
Being able to quickly update the result of a program analysis for a frequently-changed program is important for many software engineering tasks, i.e., inside an intelligent IDE or during the continuous integration (CI) / continuous development (CD) process. 
Computing the analysis result for the changed program from scratch is not only time-consuming but also wasteful when the change is small. In contrast, incrementally updating the analysis result by leveraging the intermediate results computed for a previous version of the program can be significantly faster.
While the potential for APA to support incremental computation has been mentioned before, e.g., by Kincaid et al.~\cite{kincaid2021algebraic}, exactly how to accomplish it remains unknown.

At the most fundamental level, there are two technical hurdles associated with incrementally conducting APA.  The first one is designing data structures and algorithms that can efficiently update the path expression.  The second one is designing data structures and algorithms that can efficiently updates the program properties (facts) by interpreting the changed path expression incrementally.  
Recall that in the context of APA, the path expression is a special type of regular expression for capturing  the set of all program paths of interest.  Classic APA methods focus primarily on optimizing the data structures and algorithms to compute the path expression quickly, e.g., using graph-theoretic techniques that combine tree decompositions and centroid decompositions~\cite{tarjan1981fast,conrado2023exploiting}. 
There are also methods for improving the quality of path expression~\cite{cyphert2019refinement}. While all path expressions are guaranteed to capture the feasible program paths of interest, thus guaranteeing soundness, a path expression is considered better than another if it captures fewer infeasible program paths. 
\revised{%
However, all the existing APA algorithms are optimized for non-incremental applications.
}%
Unfortunately, for incremental APA, these \emph{otherwise-elegant} optimization techniques may become a hurdle for supporting efficient updates in response to frequent program changes.  
\revised{%
Existing APA algorithms do not guarantee that small code changes lead to small incremental updates. 
}%
For example, even if a program slightly changes, classic data structures for representing path expression may change drastically. 
Similarly, classic algorithms for interpreting the path expression are not optimized to support efficient updates in response to frequent program changes.

To overcome the aforementioned limitations, we propose a new method specifically designed for incremental APA.  The goal is to drastically reduce the analysis time by leveraging intermediate results that have already been computed for a previous version of the program.  At a high level, these intermediate results include both data structures and algorithms for representing and updating path expression and program facts computed using the path expression.   

\revised{
Following classic results~\cite{RamalingamR96} on the computational complexity of dynamic graph problems, efficiently maintaining a dynamic graph’s structure and node/edge information for incremental queries is a theoretically challenging problem in general. For APA, coming up with a desirable data structure is a non-trivial task. We accomplish this by starting from a directed acyclic graph, then adding backward edges to form the basic structures, and finally designing algorithms to keep the tree balanced to efficiently handle queries. Specifically, we have chosen a “weight balanced tree” where  structural limitations imposed by the “choice” and “Kleene star” operators of APA do not allow arbitrarily rotating the tree (an operation allowed by other balanced trees such as Splay Tree or Red-Black Tree). In addition to avoiding the rotation of the tree, we also support persistent and revocable operations.
}

\begin{figure}
\vspace{2mm}
\centering

\includegraphics[width=\linewidth]{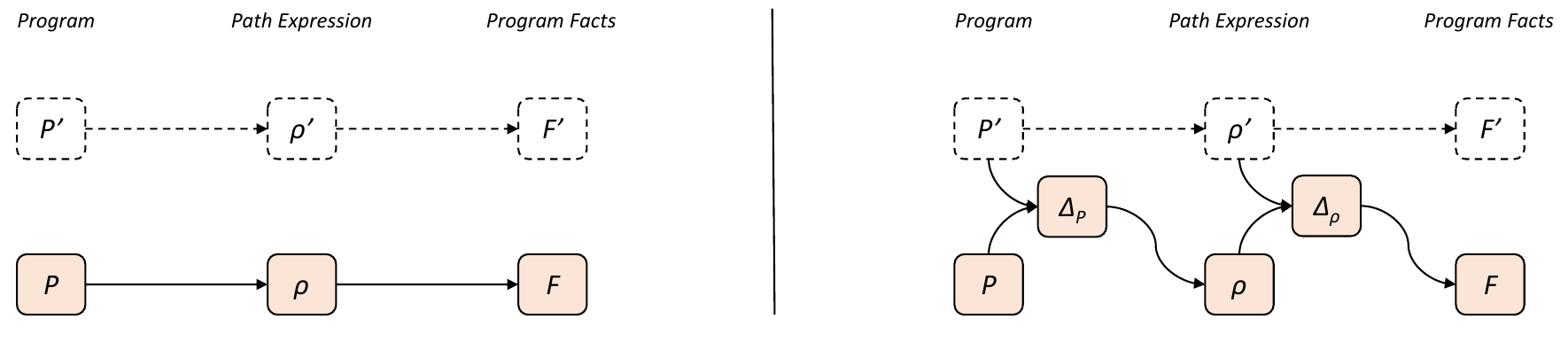}

\caption{The difference between the baseline APA (on the left) and our new incremental APA (on the right).}
\label{fig:diagram}
\end{figure}

As shown in Fig.~\ref{fig:diagram}, assuming that APA has been conducted for a previous version of the program $P'$ to obtain the path expression $\rho'$ and the set $F'$ of program facts, our goal is to efficiently compute, for the new program $P$, the path expression $\rho$ and the set $F$ of program facts. 
Instead of computing $\rho$ and $F$ for program $P$ from scratch, as shown by the baseline APA on the left-hand side, we propose to do so incrementally as shown by the new method on the right-hand side.  
\revised{Our key observation is that, given a semantic algebra, the denotational semantics of the program also satisfies the algebraic rules, which inspired us to find an efficient way to reuse the interpretation result of the prior path expression to speed up the interpretation of the new path expression.}
Specifically, we compute the new path expression $\rho$ by incrementally updating the existing path expression $\rho'$ based on the difference between programs $P'$ and $P$, denoted $\Delta_P=\textsc{Diff}(P,P')$. 
Then, we compute the new set $F$ of program facts by incrementally updating the existing set $F'$ based on $\Delta_\rho =\textsc{Diff}(\rho',\rho)$, which is the difference between path expressions $\rho'$ and $\rho$.

In practice, incrementally computing path expression $\rho$ and the program facts in $F$ based on the existing $\rho'$ and $F'$ has the potential to achieve orders-of-magnitude speedup compared to computing $\rho$ and $F$ from scratch, especially for small program changes. 
For example, with 4\% of program change, we have observed a speed up of more than  160$\times$ to 4761$\times$ during our experimental evaluation (Section~\ref{sec:experiment}), depending on the types of analysis performed by APA\footnote{In our experiments, we have implemented three types of analyses: computing reaching definitions, the use of possibily-unitialized variables, and a simple constant-time analysis for proving the absence of timing side-channel.}.
Theoretically, incrementally computing path expression $\rho$ and program facts in $F$ using APA  can also lead to some nice properties; for example, when the semantic algebra satisfies certain conditions, our method guarantees to return unique analysis result efficiently. Details can be found in Section~\ref{sec:discussion}.

While there is a large body of existing work on APA, our method differs in that it solves a significantly different problem.  
For example, the most recent work of Conrado et al.~\cite{conrado2023exploiting} focuses on amortizing the cost of answering a large number of APA queries for a fixed program (without any program change) by precomputing intermediate results while building the path expression, to achieve the goal of answering each APA query in $O(k)$ time, where $k$ is the time needed to evaluate an atomic operation in the semantic algebra. Their method exploits sparseness of the control flow graph in a centroid-based divide-and-conquer algorithm for computing path expression. 
Another recent work of Cyphert et al.~\cite{cyphert2019refinement} focuses on computing a path expression that is of a higher quality than a given path expression, assuming that both capture all feasible program paths but one is more accurate in that it captures fewer infeasible program paths. 
The classic APA methods of Tarjan, which focus on quickly computing path expression~\cite{tarjan1981fast} and a unified framework for solving path problems~\cite{tarjan1981unified}, remain competitive in terms of speed;  Reps et al.~\cite{reps1995precise} are the first to leverage  Tarjan's algorithm to compute path expression in polynomial time.  
While all of these existing methods are closely related, they do not solve the same problem as ours. Thus, we consider them to be  orthogonal and complementary to our method.

To evaluate the performance of our method in practice, we have implemented our method in a tool for analyzing Java bytecode programs, and evaluated it on 13 real-world applications from the DaCapo benchmark suite~\cite{DaCapo:paper}.  They are open-source applications implementing a diverse set of functionalities, with code size ranging from 23k LoC (\texttt{antlr}) to 220k LoC (\texttt{fop}). 
We experimentally compared our incremental APA method with three other methods: the baseline APA, the most recent method of Conrado et al.~\cite{conrado2023exploiting}, and the fast algorithm of Tarjan~\cite{tarjan1981fast}.  Our experiments were conducted using semantic algebras and semantic functions designed for three types of program analyses: computing reaching definitions, computing the use of possibly-uninitialized variables, and constant-time analysis. 
Our ablation studies show that both our technique for incrementally computing path expression and our technique for incrementally interpreting path expression are effective in speeding up the analysis. Overall, the speedup of our method is more than 160X to 4761X compared to the baseline APA and the other two existing methods.

To summarize, this paper makes the following contributions:
\begin{itemize}
\item We propose a method for conducting APA incrementally in response to program changes, with the goal of leveraging intermediate computation results to speed up the analysis. 
\item We propose new data structures and  algorithms for efficiently updating the path expression, and interpreting the path expression to compute the program facts. 
\item We implement the proposed method in a software tool and demonstrate its advantage over competing methods on Java applications from a well-known benchmark suite.
\end{itemize}

The remainder of this paper is organized as follows.  First, we provide the technical background in Section~\ref{sec:background}.  Then, we define the incremental APA problem and present our top-level procedure in Section~\ref{sec:method_overall}.  Next, we present our techniques for incrementally computing the path expression in Section~\ref{sec:method_compute}, and for incrementally interpreting the path expression in Section~\ref{sec:method_interpret}. We analyze the mathematical properties of our proposed method in Section~\ref{sec:discussion}, and present the experimental results in Section~\ref{sec:experiment}. We review the related work in Section~\ref{sec:related}, and finally, give our conclusion in Section~\ref{sec:conclusion}.

\section{Background}
\label{sec:background}

In this section, we review the basics of algebraic program analysis (APA) and present  the technical details of the two distinct steps of APA: computing the path expression and interpreting the path expression to compute the program facts of interest.

\subsection{The Program}

Given a program $P$, algebraic program analysis is concerned with computing \emph{facts} that must be true at each program location, regardless of the actual path of program execution taken to reach the location. 
To solve the problem, a preliminary step is constructing the control flow graph, denoted $G = (N, E, s)$ where $N$ is a set of nodes, $E$ is a set of directed edges between the nodes, and $s\in N$ is the entry node.  
Each node $n\in N$ represents a \emph{basic block} of the program, where a basic block is a block of contiguous program statements with a single entry and a single exit.  
Each edge $e\in E$ represents a possible transfer of control between the nodes (basic blocks) in $N$.  The entry node $s\in N$ represents the start of program execution.  

\begin{figure}
\vspace{2mm}
\centering
\begin{minipage}{0.33\linewidth}
\centering
\begin{lstlisting}[language=C,firstnumber=1]
    int a, b, c, d, e;
    a = 5;
    b = a + 5;
    while (a < 20){
        
        if (a > 0) 
            d = b;
        else a = a + 10;
        printf(''\%d'', d);
    }
    c = a + 1;
    printf(''\%d'', b+c+e);
    return;
\end{lstlisting}
\end{minipage}
\hspace{0.05\linewidth}
\begin{minipage}{0.6\linewidth}
    \centering
    \includegraphics[width=0.9\linewidth]{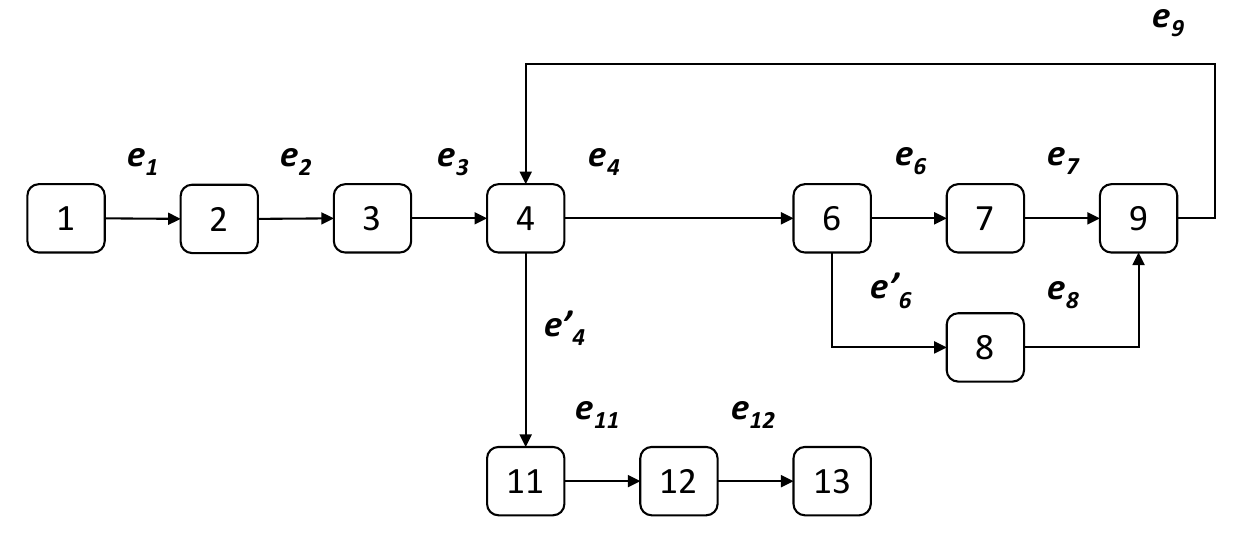}
\vspace{2ex}
\end{minipage}
\caption{An example program for the problem of detecting uses of possibly-uninitialized variables.} 
\label{fig:ex-program}
\end{figure}

Consider the example program shown on the left-hand side of Fig.~\ref{fig:ex-program}, for which the control flow graph is shown on the right-hand side. Assuming that each line in the program corresponds to a node in the graph, we have the set of nodes $N=\{n_1-n_4, n_6-n_9, n_{11}-n_{13}\}$, the set of edges $E=\{e_1-e_3,e_4,e'_4,e_6,e'_6,e_7-e_9,e_{11},e_{12}\}$ and the entry node $s = n_1$. Specifically, the two edges coming out of node $n_4$ are labeled $e_4$ and $e'_4$, respectively, and the two edges coming out of node $n_6$ are labeled $e_6$ and $e'_6$, respectively.

\subsection{Two Approaches to Program Analysis}

As mentioned earlier, there are two approaches to program analysis: iterative program analysis and algebraic program analysis.  
Iterative program analysis starts with a system of (recursive) equations for defining the semantics of the program in an abstract domain, followed by solving these equations through successive approximation. In the literature, the first step is referred to as \emph{interpret} and the second step is referred to as \emph{compute}.   Thus, iterative program analysis is also called the \emph{interpret-and-then-compute} approach. Examples of iterative program analysis include Kildall's gen/kill analysis~\cite{Kildall73},  the abstract interpretation framework~\cite{CousotC77}, and model checking of Boolean programs using predicate abstraction~\cite{BallMMR01}.

Algebraic program analysis, in contrast, adopts the \emph{compute-and-then-interpret} approach.  It starts with computing the path expression, which is a type of regular expression for summarizing the program paths of interest, followed by interpreting the path expression using a properly-defined semantic algebra and a semantic function, to obtain the program facts of interest.  
Regardless of which type of analysis is performed by APA, the path expression remains the same for a given program. However, the definitions of semantic algebra and semantic function will have to depend on the nature of the analysis itself.  
Depending on the task, the semantic algebra and semantic function must be defined accordingly.
Nevertheless, at a high level, all types of analyses performed by APA have the same two steps: computing the path expression and interpreting the path expression.

\subsection{The Path Expression}

Given a control flow graph $G = (N,E,s)$, the set of all program paths that start from $s$ and end at $t$ may be captured by a special type of regular expression over $E$, denoted $\rho(s,t)$, \revised{where $s$ is the start point of a path, and $t$ is the end point}.  This regular expression is referred to as the \emph{path expression}~\cite{tarjan1981fast,tarjan1981unified}. 
We say that the regular expression is \emph{special} because the finite alphabet used to define the regular expression is $E$, the set of edges in the control flow graph.  Thus, a string corresponds to a sequence of edges in the control flow graph.  While not every string corresponds to a program path, e.g., it may have edges randomly scattered in the graph, not forming a path at all,  we require that all strings in the path expression to correspond to  program paths.

The path expression $\rho$ is defined recursively as follows: 
\begin{itemize}
\item $\epsilon$ (the empty string) is an atomic regular expression.
\item $\emptyset$ (the empty set) is an atomic regular expressions.
\item Any edge $e\in E$ is an atomic regular expression.
\item Given two regular expressions $\rho_1$ and $\rho_2$, the union ($\rho_1 + \rho_2$), concatenation ($\rho_1\cdot \rho_2$) and Kleene star $(\rho_1)^*$ are regular expressions. 
\end{itemize}
In addition, if a regular expression $\rho_1$ is repeated $k$ times, we represent it using $\rho_1^k$. That is, $\rho_1^k=\rho_1^{k-1}\cdot \rho_1$, where $k\geq 1$ and $\rho_1^0 = \epsilon$ (the empty string).

Consider the program shown in Fig.~\ref{fig:ex-program} as an example.  The path expression that captures the first three lines of code is $\rho_1=e_1e_2e_3$, which is the concatenation of the three edges.   The path expression that captures Lines 6-8 is $\rho_2=(e_6e_7 + e'_6e_8)$, which is the union of the two branches.  The path expression that captures the while-loop is $\rho_3=(e_4 (e_6e_7 + e'_6e_8) e_9)^*$. 
Finally, the path expression that captures all paths from $n_1$ to $n_{13}$ is 
$\rho = e_1e_2e_3(e_4 (e_6e_7 + e'_6e_8) e_9)^* e'_4e_{11}e_{12}$, where $e_1e_2e_3$ represents the prefix leading to the while-loop, $(e_4 (e_6e_7 + e'_6e_8) e_9)^*$ represents the while-loop, and $e'_4e_{11}e_{12}$ represents the suffix after the while-loop.

At this moment, it is worth noting that the path expression guarantees to capture all the \emph{feasible} program paths, thus leading to guaranteed soundness of APA.  At the same time, not all program paths captured by the path expression may be feasible, meaning that APA (same as iterative program analysis) in general is a type of possibly conservative (over-approximate) analysis.   In other words, APA is a sound, but not-necessarily-complete analysis.

\subsection{The Semantic Algebra}
\label{sec:semalgebra}

Given a path expression $\rho$, which summarizes the program paths of interest, we can compute the program facts over these program paths, by interpreting the path expression using a properly defined semantic algebra, denoted  $\mathcal{D} =\langle D, \otimes,\oplus, \circledast, \mathtt{0}, \mathtt{1}\rangle$.  Here, 
$D$ is the universe of program facts, 
$\otimes: D\times D\rightarrow D$ is the sequencing operator,
$\oplus: D\times D\rightarrow D$ is the choice operator, and
$\circledast: D\rightarrow D$ is the iteration operator.  Furthermore, 
$\mathtt{0}$ and $\mathtt{1}$  are the minimal and maximal elements in $D$, respectively.

Intuitively, the $\otimes$, $\oplus$ and $\circledast$ operators in the semantic algebra $\mathcal{D}$ correspond to the concatenation ($\times $), union ($+$) and Kleene star ($*$) operators in regular expression, respectively.  
Thus, we have 
\begin{itemize}
\item
$\mathcal{D} \llbracket \rho_1\rho_2 \rrbracket = \mathcal{D}\llbracket \rho_1\rrbracket \otimes
                                           \mathcal{D}\llbracket \rho_2\rrbracket$;
\item
$\mathcal{D} \llbracket \rho_1+\rho_2 \rrbracket = \mathcal{D}\llbracket \rho_1 \rrbracket \oplus
                                           \mathcal{D}\llbracket \rho_2 \rrbracket$; and 
\item
$\mathcal{D} \llbracket (\rho_1)^* \rrbracket = (\mathcal{D}\llbracket \rho_1\rrbracket)^{\circledast}
$ 
.
\end{itemize}
What it means is that program facts can be computed in a bottom-up fashion, first for small components, $\rho_1$ and $\rho_2$, and then for the large component $\rho=\rho_1\rho_2$, $\rho=\rho_1+\rho_2$, or $\rho=(\rho_1)^*$.

\subsection{The Semantic Function}

At this moment, we have not yet defined $D$, the universe of program facts, or the actual functions for $\otimes$, $\oplus$ and $\circledast$.  The reason is because they must be defined for each type of analysis, whether it is for computing reaching definitions or for computing the use of possibly-uninitialized variables.  
For ease of understanding, in the remainder of this section, we treat the use of possibly-uninitialized variables as a running example.

\myparagraph{Running Example: Use of Possibly-uninitialized Variables}
Let $Var$ be the set of  variables in a program $P$.  Given a subset of $Var$, each variable $v\in Var$ is either in that subset, or outside of that subset; in other words, there are only 2 possible cases.  Thus, the power-set $2^{Var}$ consists of all possible subsets of $Var$.  
To capture the space of program facts,  we define $D= (DI,PU)$ where 
\begin{itemize}
\item $DI= 2^{Var}$ stands for the \emph{definitely-initialized} set,  consisting of all possible subsets of variables that are definitely-initialized, and 
\item $PU=2^{Var}$ stands for the \emph{possibly-uninitialized} set, consisting of all possible subsets of variables that are used while being possibly-uninitialized. 
\end{itemize}

After defining the space of program fact $D=(DI,PU)$, for each edge $e\in E$,
we define a semantic function $\mathcal{D}\llbracket  \rrbracket: E\rightarrow D$.  For brevity, we only show what it looks like using two assignment statements from Fig.~\ref{fig:ex-program}: 
\begin{itemize}
\item for the edge $e_2$ coming out of node $n_2$: \texttt{a=5}, we define $DI_{e_2}=\{a\}$ and $PU_{e_2}=\{\}$;
\item for the edge $e_3$ coming out of node $n_3:$ \texttt{b=a+5}, we define $DI_{e_3}=\{b\}$ and $PU_{e_3}=\{a\}$.
\end{itemize}
The reason why $DI_{e_3}=\{b\}$, meaning $b$ is \emph{definitely-initialized}, is because the statement in $n_3$ writes to $b$.  The reason why $PU_{e_3}=\{a\}$ is because the statement in $n_3$ reads from $a$, but $a$ is not yet defined in $n_3$ alone; thus, we assume it is a \emph{use of possibly-uninitialized variable} for now.

For the $\otimes$ operator, e.g., $L\otimes R$,
we define a semantic function. 
%
Assuming that $D_L=(DI_L,PU_L)$ and $D_R=(DI_R,PU_R)$ are already computed, we define $D_{L\otimes R}=(DI,PU)$ as follows: 
      \begin{itemize}
      \item  $DI = DI_L \cup DI_R$; and 
      \item  $PU = PU_L \cup (PU_R \setminus DI_L)$. 
      \end{itemize}
Consider $e_2\otimes e_3$ from the program in Fig.~\ref{fig:ex-program} as an example.  The reason why $DI = DI_{e_2} \cup DI_{e_3} = \{a,b\}$ is because, as long as a variable is definitely-initialized in either $e_2$ or $e_3$, it is definitely-initialized in $e_2\otimes e_3$. 
The reason why $PU = PU_{e_2} \cup (PU_{e_3}\setminus DI_{e_2}) = \{ \}$ is because, although $a$ is used while being  not-yet-initialized in $e_3$, it is initialized in $e_2$; therefore, $a$ is removed from the set $PU$.

For the $\oplus$ operator, e.g., $L\oplus R$,
%
we define  $D_{L\oplus R}=(DI,PU)$ as follows:
      \begin{itemize}
      \item  $DI = DI_L \cap DI_R$; and 
      \item  $PU = PU_L \cup PU_R$. 
      \end{itemize}
Consider $(e_6\otimes e_7 \oplus e'_6\otimes e_8)$ from the program in Fig.~\ref{fig:ex-program} as an example.  The reason why $DI = DI_{e_6\otimes e_7} \cap DI_{e'_6\otimes e_8} = \{d \}\cap \{a \} =\{\}$ is because a variable is definitely-initialized if it is definitely-initialized in both branches.
The reason why $PU = PU_{e_6\otimes e_7} \cup PU_{e'_6\otimes e_8} = \{a,b \}\cup \{a \} =\{a,b\}$ is because a variable is possibly-uninitialized if it is possibly-uninitialized in either of the branches.

For the $\circledast$ operator, e.g., $(L)^\circledast$,
%
we define $D=(DI,PU)$ as follows: 
      \begin{itemize}
      \item  $DI = \emptyset$; and 
      \item  $PU = PU_L$.
      \end{itemize}
The reason  why $DI=\emptyset$ is because $(L)^\circledast$ includes $(L)^0=\epsilon$, which is an empty string representing the skip of the loop body; in this case, no variable is defined at all.   
The reason why $PU=PU_L$ is because, if a variable is used while being possibly-uninitialized during one loop iteration, it remains a use of possibly-uninitialized variable for an arbitrary number of loop iterations. 

\subsection{The Baseline APA}

In the remainder of this section, we explain how the baseline (non-incremental) APA works using the program in Fig.~\ref{fig:ex-program} as the running example. 
Our goal is to compute the use of possibly-uninitialized variables. 
In this program, there are two such uses.  The first one is the use of variable \texttt{d} at Line~9, where \texttt{d} may be uninitialized if the else-branch at Line~8 is executed.  The second one is the use of variable \texttt{e} at Line~12, which is not initialized in this program.  In both cases, the variables are possibly-uninitialized when they are used.\footnote{If a more accurate analysis is preferred, these two uses may be separated further: while \texttt{d} at Line~9 \emph{may} be uninitialized, \texttt{e} at Line~12 \emph{must} be uninitialized.}

As mentioned earlier, APA consists of two distinct steps: computing the path expression and interpreting the path expression. In the baseline (non-incremental) APA, both steps are performed in a bottom-up fashion, first for individual edges in $E$, then for small code fragments, and finally for the entire program.

\myparagraph{Computing the Path Expression}
The first step of APA is to compute the path expression, which summarizes the program paths of interest.   While there can be many data structures for representing a regular expression and all of them have the same semantics (meaning they capture the same set of strings), for the purpose of representing the path expression in APA, they may behave differently. The reason is because the chosen data structure may significantly affect both the speed and the result of the analysis. 
Depending on the order in which subexpressions are evaluated, for example, the running time of APA  may be different. 
Furthermore, the analysis result (the set of program facts computed by APA at the end) may also be different.

\begin{figure}
\vspace{2mm}
\centering
    \includegraphics[width=0.8\linewidth]{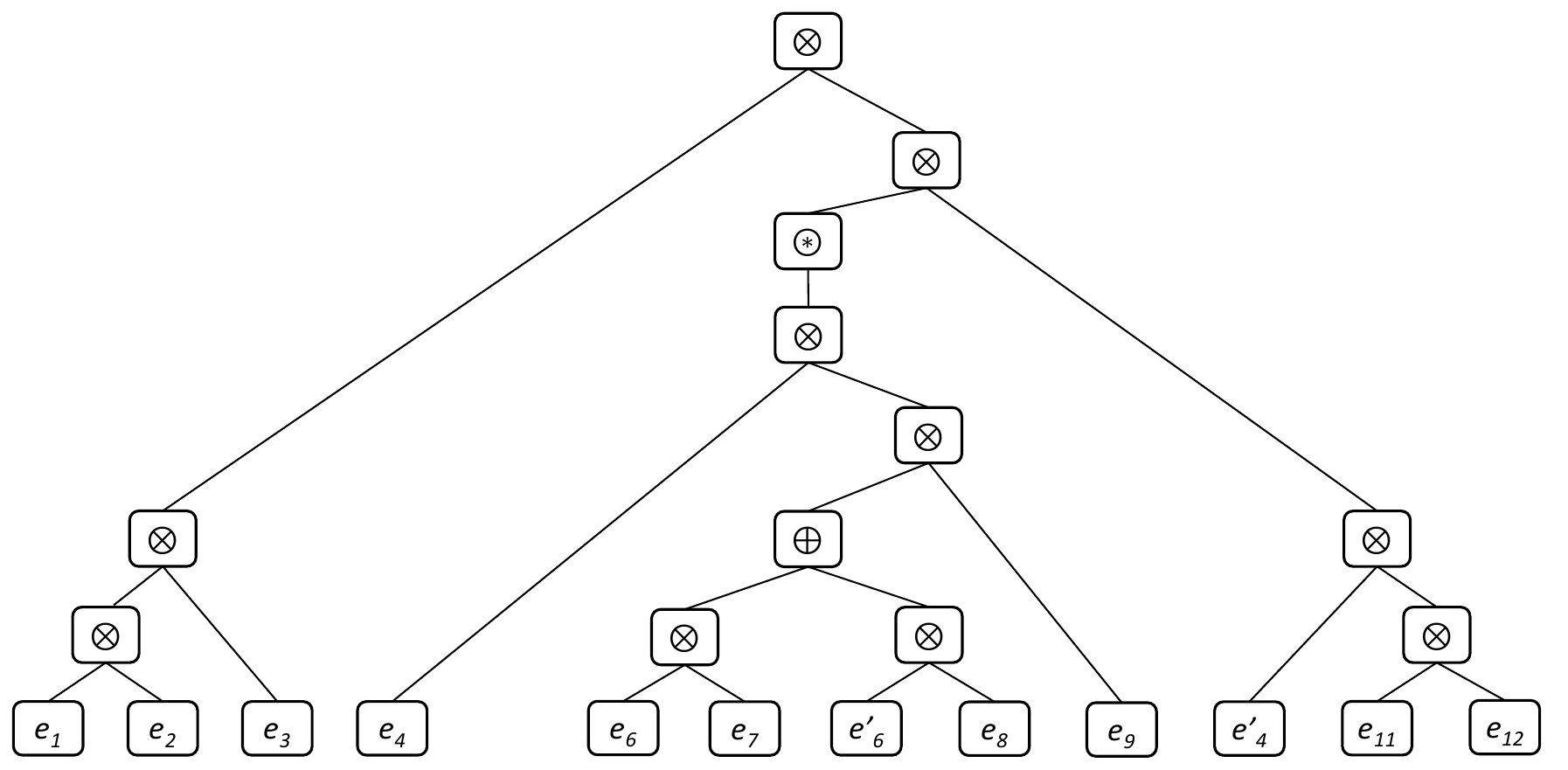}
\caption{Computing the program properties by interpreting the path expressions. }
\label{fig:ex-tree}
\end{figure}

For the analysis result to be unique, the semantic algebra must be \emph{associative}.  
In Section~\ref{sec:discussion}, we will discuss the impact of various algebraic properties in detail. 
For now, it suffices to assume that the path expression will be represented using a tree, as shown in Fig.~\ref{fig:ex-tree}.

In this figure, the terminal nodes correspond to edges in the control flow graph.  The internal nodes correspond to the sequencing ($\otimes$), choice ($\oplus$), and iteration ($\circledast$) operators.   
For example, the $\otimes$ node that combines $e_1$ and $e_2$ represents the subexpression $e_1 e_2$, meant  to be interpreted as $e_1\otimes e_2$.  
The $\oplus$ node that combines the nodes $e_6e_7$ and $e'_6e_8$ represents the subexpression $(e_6e_7 + e'_6e_8)$, meant to be interpreted as $(e_6\otimes e_7 \oplus e'_6\otimes e_8)$.
The while-loop is represented by $(e_4 (e_6e_7 + e'_6e_8) e_9)^*$, meant to be interpreted as $(e_4 \otimes (e_6\otimes e_7 \oplus e'_6\otimes e_8)\otimes e_9)^\circledast$.
Finally, the entire program is represented by $e_1e_2e_3(e_4 (e_6e_7 + e'_6e_8) e_9)^* e'_4e_{11}e_{12}$, meant to be interpreted as $e_1\otimes e_2\otimes e_3\otimes (e_4 \otimes (e_6\otimes e_7 \oplus e'_6\otimes e_8)\otimes e_9)^\circledast \otimes e'_4\otimes e_{11}\otimes e_{12}$.

\myparagraph{Interpreting the Path Expression}
The second step of APA is to interpret the path expression according to the semantic algebra and semantic function.  For computing the use of possibly-uninitialized variables, we shall use the semantic algebra and semantic function defined in the previous subsection for illustration purposes.
For the path expression tree shown in Fig.~\ref{fig:ex-tree}, interpretation is performed in a bottom-up fashion.  That is, we first compute $D_L=(DI_L,PU_L)$ for subexpression $L$  and $D_R=(DI_R,PU_R)$ for subexpression $R$, and then compute $D_{L\otimes R}$, $D_{L\oplus R}$ and $D_{L^\circledast}$ for the compound expressions.

The table below shows the detailed process of interpreting the path expression, to obtain the program facts in $D=(DI,PU)$ for the entire program:

\vspace{1ex}

\noindent
\scalebox{0.98}{\scriptsize
\begin{tabular}{|l|l|l|}
\hline
Path expression &  Definitely-initialized variables  & Use of possibly-uninitialized variables \\[0.5ex]\hline\hline
$e_1$     &    $DI_{e_1}=\{   \}$ & $PU_{e_1} =\{   \}$ \\[0.5ex]
$e_2$     &    $DI_{e_2}=\{ a \}$ & $PU_{e_2} =\{   \}$ \\[0.5ex]
$e_1e_2$  &    $DI_{e_1e_2} = DI_{e_1} \cup DI_{e_2} = \{ a \}$ &  
               $PU_{e_1e_2} = PU_{e_1} \cup (PU_{e_2} \setminus DI_{e_1}) = \{ \}$\\[0.5ex]
$e_3$     &    $DI_{e_3}=\{ b \}$ & $PU_{e_3} =\{ a \}$\\[0.5ex]
$e_1e_2e_3$ &    $DI_{e_1e_2e_3} = DI_{e_1e_2} \cup DI_{e_3} = \{ a,b \}$ & 
               $PU_{e_1e_2e_3} = PU_{e_1e_2} \cup (PU_{e_3}\setminus DI_{e_1e_2}) = \{ \}$\\[0.5ex]
$e_4$     &    $DI_{e_4}=\{ \}$ & $PU_{e_4} =\{ a \}$ \\[0.5ex]
$e_6$     &    $DI_{e_6}=\{ \}$ & $PU_{e_6} =\{ a \}$ \\[0.5ex]
$e_7$     &    $DI_{e_7}=\{ d \}$ & $PU_{e_7}=\{ b \}$ \\[0.5ex]
$e_6e_7$  &    $DI_{e_6e_7}=\{ d \}$ & $PU_{e_7}=\{ a,b \}$ \\[0.5ex]
$e_8$     &     $DI_{e_8}=\{ a \}$ & $PU_{e_8}=\{ a \}$ \\[0.5ex]
$e_6'e_8$  &     $DI_{e'_6e_8}=\{ a \}$ & $PU_{e'_6e_8}=\{ a \}$ \\[0.5ex]
$(e_6e_7+e'_6e_8)$    &   $DI_{(e_6e_7+e'_6e_8)}= \{  \}$ & $PU_{(e_6e_7+e'_6e_8)}= \{ a,b \}$ \\[0.5ex]
$e_9$     &     $DI_{e_9}=\{  \}$ & $PU_{e_9}=\{ d \}$ \\[0.5ex]
$(e_6e_7+e'_6e_8)e_9$ &   $DI_{(e_6e_7+e'_6e_8)e_9}= \{  \}$ & $PU_{(e_6e_7+e'_6e_8)e_9}= \{ a,b,d \}$ \\[0.5ex]
$e_4(e_6e_7+e'_6e_8)e_9$ &   $DI_{e_4(e_6e_7+e'_6e_8)e_9}= \{  \}$ & $PU_{e_4(e_6e_7+e'_6e_8)e_9}= \{ a,b,d \}$ \\[0.5ex]
$(e_4(e_6e_7+e'_6e_8)e_9)^*$ &  $DI_{(e_4(e_6e_7+e'_6e_8)e_9)*}=\{  \}$ & $PU_{(e_4(e_6e_7+e'_6e_8)e_9)*}=\{ a,b,d \}$ \\[0.5ex]
$e_{11}$  &    $DI_{e_{11}} = \{ c \}$ & $PU_{e_{11}} = \{a\}$ \\[0.5ex]
$e_{12}$  &    $DI_{e_{12}} = \{ \}$   & $PU_{e_{12}} = \{b, c, e\}$ \\[0.5ex]
$e_{11}e_{12}$  & $DI_{e_{11}e_{12}} = \{ c\}$ & $PU_{e_{11}e_{12}} = \{a,b,e\}$ \\[0.5ex]
$(e_4(e_6e_7+e'_6e_8)e_9)^*e'_4e_{11}e_{12}$   & $DI_{(e_4(e_6e_7+e'_6e_8)e_9)^*e_{11}e_{12}} = \{c\}$ & $PU_{(e_4(e_6e_7+e'_6e_8)e_9)^*e'_4e_{11}e_{12}} = \{ a, b, d, e \}$ \\[0.5ex]\hline\hline
$e_1e_2e_3(e_4(e_6e_7+e'_6e_8)e_9)^*e'_4e_{11}e_{12}$   & $DI_{e_1e_2e_3(e_4(e_6e_7+e'_6e_8)e_9)^*e'_4e_{11}e_{12}} =  \{a,b,c\}$ &
                $PU_{e_1e_2e_3(e_4(e_6e_7+e'_6e_8)e_9)^*e'_4e_{11}e_{12}} = \{d,e\}$ \\[0.5ex]\hline
\end{tabular}
}

\vspace{1ex}

\noindent
For example, consider the last row of the above table, which takes the program facts computed for $\rho_1 = e_1e_2e_3$ and $\rho_2=(e_4(e_6e_7+e'_6e_8)e_9)^*e'_4e_{11}e_{12}$ as input, and returns the program facts for the entire program ($\rho$) as output. 
At the start of the computation, we have $DI_{\rho_1} = \{ a, b\}$ and $PU_{\rho_1} = \{ \}$, meaning $a$ and $b$ are definitely-initialized in $\rho_1$.
We also have $DI_{\rho_2} = \{ c \}$ and $PU_{\rho_2} = \{ a,b,d,e \}$, meaning that $c$ is definitely-initialized in $\rho_2$ and $a,b,c,e$ are the uses of possibly-uninitialized variables in $\rho_2$ alone.

Since $\rho = \rho_1\otimes\rho_2$, we have 
$DI_{\rho} = DI_{\rho_1} \cup DI_{\rho_2} = \{a, b, c\}$, meaning that $a,b,c$ are the definitely-initialized variables for the entire program,   and
$PU_{\rho} = PU_{\rho_1} \cup (PU_{\rho_2} \setminus DI_{\rho_1}) = \{ \} \cup (\{a,b,d,e\} \setminus \{a,b\}) = \{d,e\}$,  meaning that only $d$ and $e$ are uses of possibly-uninitialized variables; these two variables are used at Lines 9 and 12, respectively.  
In contrast, $a$ and $b$ in $PU_{\rho_2}$ are removed from $PU_{\rho}$ because these two variables are initialized in $\rho_1$, as indicated by $DI_{\rho_1}$.

\section{Our Method}
\label{sec:method_overall}

In this section, we present an overview of our method for incremental APA, which has two main components.  The first component consists of techniques for incrementally updating the path expression, to respond to changes of the program. The second component consists of techniques for incrementally interpreting the path expression, to obtain the program facts of interest. 

\begin{algorithm}
\caption{$F\leftarrow$ \textsc{Incremental\_APA}($P,P',\rho',F')$)}
\label{alg:incremental-APA}
{\footnotesize
\begin{algorithmic}
\State $\Delta_P    \leftarrow$ \textsc{Diff} ($P,P'$)
\State $\rho        \leftarrow$ \textsc{Compute\_PathExpression\_Inc}  ($\Delta_P, \rho'$)
\State $\Delta_\rho \leftarrow$ \textsc{Diff} ($\rho,\rho'$)
\State $F          \leftarrow$ \textsc{Interpret\_PathExpression\_Inc} ($\Delta_\rho, F'$)
\State \Return $F$
\end{algorithmic}
}
\end{algorithm}

Algorithm~\ref{alg:incremental-APA} shows the top-level procedure of our method. In addition to $P$, the new program, the input of our procedure includes $P'$, a previous version of the program, together with its path expression $\rho'$ and the set $F'$ of program facts.  The output of our procedure is $F$, the set of program facts for $P$.
Internally, the procedure goes through two steps.  First, it computes the difference between $P$ and $P'$, denoted $\Delta_P$, and leverages it to update $\rho'$ to obtain $\rho$. 
Next, it computes the difference between $\rho$ and $\rho'$, denoted $\Delta_\rho$, and leverages it to update $F'$ to obtain $F$.

When $\Delta_P$ is small, our goal is to keep $\Delta_\rho$ small as well.  We accomplish this by using a carefully-designed data structure for representing $\rho$ and carefully-designed techniques for updating it. 
Similarly, when $\Delta_\rho$ is small, our goal is to keep the change from $F'$ to $F$ small as well.   
Our techniques for supporting these incremental updates will be presented in Sections~\ref{sec:method_compute} and \ref{sec:method_interpret}.  For now, we shall use an example to illustrate the potential of our method in speeding up the analysis.

\begin{figure}[h]
\centering
\begin{minipage}{0.33\linewidth}
\centering
\begin{lstlisting}[language=C,firstnumber=1]
    int a, b, c, d, e;
    a = 5;
    b = a + 5;
    while (a < 20){
        b = a + 5; //added
        if (a > 0) 
            d = b;
        else a = a + 10;
        printf(''\%d'', d);
    }
    c = a + 1;
    printf(''\%d'', b+c+e);
    return;
\end{lstlisting}
\end{minipage}
\hspace{0.05\linewidth}
\begin{minipage}{0.6\linewidth}
    \centering
    \includegraphics[width=0.8\linewidth]{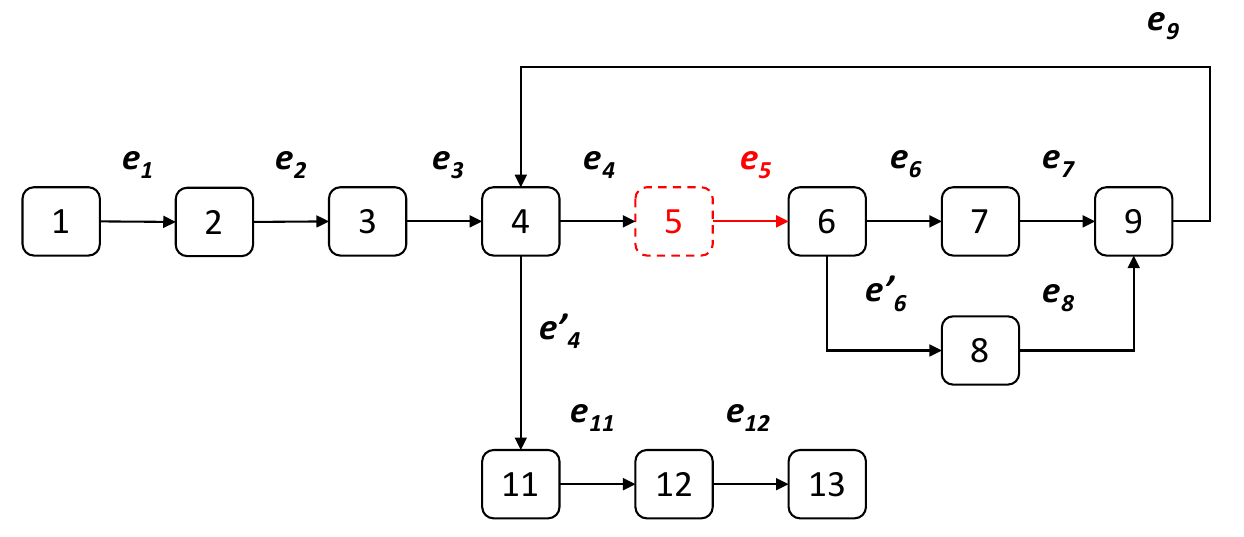}
\vspace{2ex}
\end{minipage}
\caption{The changed example program on the left-hand side and its control flow graph on the right-hand side.}
\label{fig:ex-program2}
\end{figure}

\myparagraph{The Changed Program}
Assuming that a new assignment statement, \texttt{b = a+5}, is added to Line 5 of the example program in Fig.~\ref{fig:ex-program}, which leads to the new program in Fig.~\ref{fig:ex-program2}.  For ease of presentation, we keep the line numbers of the two programs the same.   Thus, in the new control flow graph, the only change is adding node $n_5$ and edge $e_5$, both of which are highlighted in red color in Fig.~\ref{fig:ex-program2}.

\begin{figure}
\vspace{2mm}
\centering 
\includegraphics[width=0.8\linewidth]{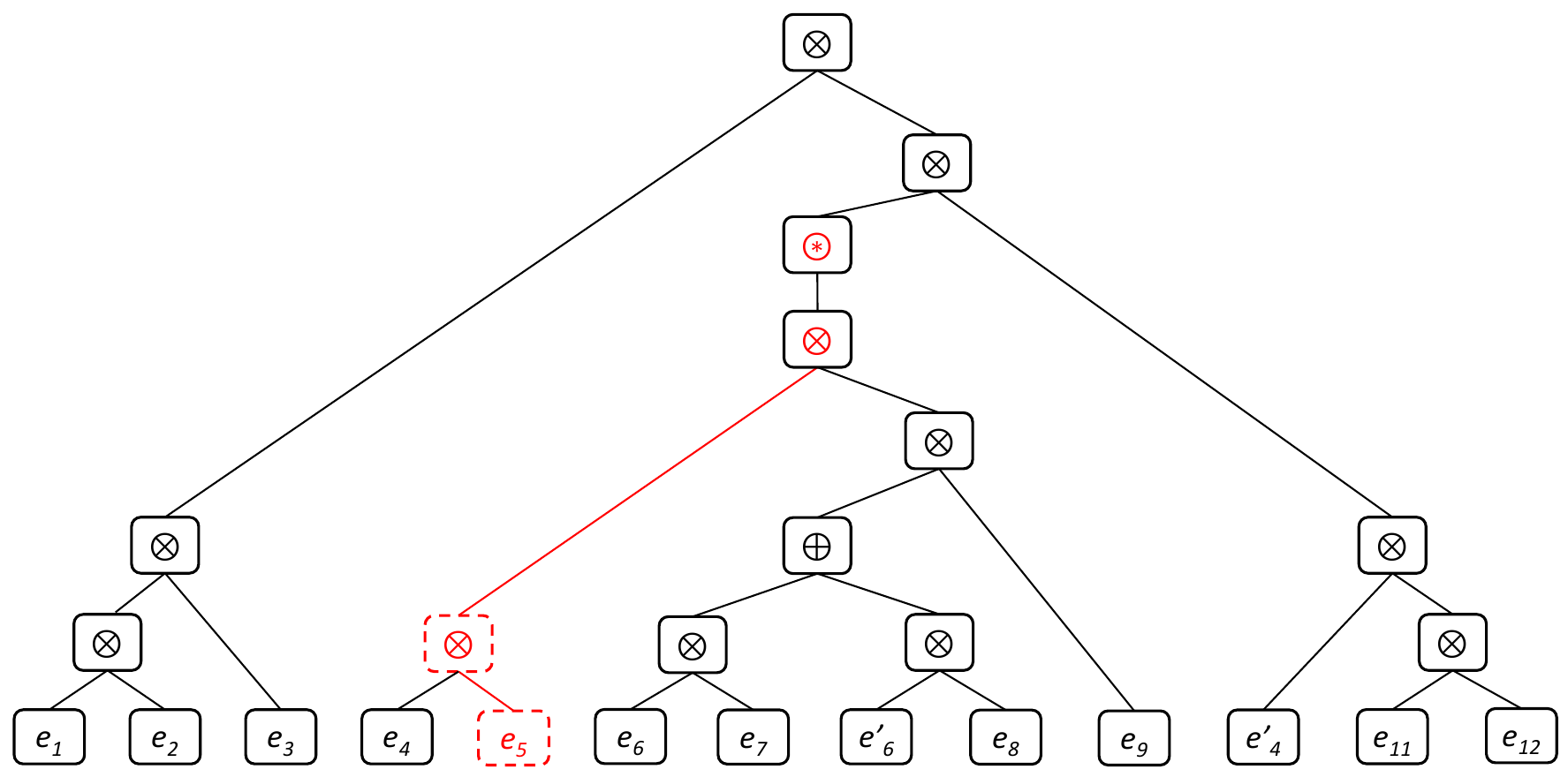}
\caption{Updating the program properties by interpreting the affected nodes in the path expression. Only two nodes need to be added; they are the nodes for $e_5$ and $e_4\oplus e_5$. Furthermore, only four nodes need to update the associated program facts; these four nodes are shown in red color.}
\label{fig:ex-tree2}
\end{figure}

\myparagraph{Updating the Path Expression}
Instead of computing the path expression $\rho$ for the new program $P$ from scratch, we choose to update the path expression $\rho'$ for the previous program $P'$.   In this context, $\rho'$ refers to  the previous path expression tree shown in Fig.~\ref{fig:ex-tree} while $\rho$ refers to the new path express tree shown in Fig.~\ref{fig:ex-tree2}.   How to carefully design the data structure and techniques to minimize the difference $\Delta_\rho = \textsc{Diff}(\rho,\rho')$ is an important research question since they may have significant effects. For example, whether the new edge $e_5$ is combined with $e_4$ first or with $e_6$ and $e'_6$ first can result in different performances. Although these two options lead to semantically-equivalent regular expressions, but the runtime performance of APA may be different: combining $e_5$ with $e_4$ first will lead to a faster APA because the tree's height will be shorter.

For the tree shown in Fig.~\ref{fig:ex-tree2}, after adding $e_5$, only one more node in the tree needs to be added;  it is the $\otimes$ operator used to combine $e_4$ and $e_5$.  In other words, the incremental update changes only 2/26 of the nodes of the path expression tree.   This is the reason why incremental update has the potential to drastically reduce the analysis time.

\myparagraph{Updating the Program Facts}
After adding $e_5$ and the $\otimes$ operator (for $e_4\otimes e_5$) to the path expression tree, we must interpret the path expression to obtain the new program facts.   Since most of the program facts computed for the previous path expression $\rho'$ remain unchanged, our method only needs to update  the program facts associated with 4/26 of the nodes in the path expression tree.  

In Fig.~\ref{fig:ex-tree2}, these four nodes (whose program facts need to be updated) are highlighted in red color. 
We show how these program facts are updated in the table below.

\vspace{2ex}
{\footnotesize
\noindent
\begin{tabular}{|l|l|l|c|}
\hline
Path expression &  Definitely-initialized variables  & Uses of possibly-uninitialized variables & Changed the facts\\\hline\hline

\textcolor{darkred}{$e_5$}       
                &    $DI_{e_5}  =\{b\}$    & $PU_{e_5}   = \{a\}$  & yes\\
\textcolor{darkred}{$e_4e_5$}  
                &    $DI_{e_4e_5}=\{b\}$    & $PU_{e_4e_5} = \{a\}$ & yes\\
\textcolor{darkred}{$e_4e_5(e_6e_7+e'_6e_8)e_9$}
                &  $DI_{e_4e_5(e_6e_7+e'_6e_8)e_9}   = \{ b     \}$ 
                &  $PU_{e_4e_5(e_6e_7+e'_6e_8)e_9}   = \{ a,b,d \}$   & yes\\
\textcolor{darkred}{$(e_4(e_6e_7+e'_6e_8)e_9)^*$}
                &  $DI_{(e_4e_5(e_6e_7+e'_6e_8)e_9)*} = \{       \}$ 
                &  $PU_{(e_4e_5(e_6e_7+e'_6e_8)e_9)*} = \{ a,b,d \}$  & no \\
\hline
\end{tabular}
}
\vspace{2ex}

This example shows that the time taken to update program facts after the addition of a terminal node in the tree may be lower than $O(h(\rho))$, where $h(\rho)$ is the height of the updated path expression tree. The reason why the complexity is $O(h(\rho))$ in general is because $h(\rho)$ is the length of the path from the root to any terminal node. 
In the above example, the height of the tree is 6, but due to early termination of the updating process, only 4 of the nodes (on the root-to-$e_5$ path) are updated.

\myparagraph{Advantages of Incremental APA}
We have illustrated the main advantage of incremental APA, which is the significantly-higher analysis  speed.  We will show, through experimental evaluation in Section~\ref{sec:experiment}, that the speedup can be more than 160$\times$ to 4761$\times$, depending on the actual type of analysis performed by APA.

Another advantage of incremental APA is that the method is inherently compositional.  That is, subexpressions in the tree may be updated in isolation, before their results are combined using algebraic rules.  It allows the method to be easily parallelized. 
For example, if the program change $\Delta_P=\texttt{Diff}(P,P')$ involves multiple root-to-leaf edges, meaning that multiple paths need to be recomputed in the tree, in addition to the red root-to-$e_5$ path in Fig.~\ref{fig:ex-tree2}, it is possible to handle these paths in parallel. 
This is an advantage of incremental APA.  Iterative program analysis does not support such parallelization because its computation is inherently sequential: every application of a transfer function depends on its input values.

\section{Incrementally Computing Path Expression} 
\label{sec:method_compute}

In this section, we first review the existing algorithms for computing path expression, and then present our new algorithm for \emph{incrementally} computing path expression. 

\subsection{The Non-incremental Algorithm}
\label{subsec3.1:ds}

Classic APA methods~\cite{tarjan1981fast,kincaid2021algebraic,conrado2023exploiting} rely on various algorithmic techniques to compute path expression efficiently.  A common theme is batch processing of the program paths of interest.  Different ways of batch processing have led to algorithms with different time complexities.  For example, Tarjan~\cite{tarjan1981fast} used dominator tree for batch processing, while Conrado et al.~\cite{conrado2023exploiting} used tree-decomposition together with a centroid-based divide-and-conquer algorithm.  However, as mentioned earlier, these algorithmic techniques are not suitable for incremental computation.

Instead, our method uses the algorithm presented by Kincaid et al.~\cite{kincaid2021algebraic} as its baseline.   Before extending the baseline for incremental computation in the next subsection, we briefly review the algorithm.   The algorithm relies on the notion of \emph{subprogram}, which is a maximal contiguous sequence of basic blocks within a program that does not contain branches or loop structures. Computing the path expression for a subprogram is easy because there is only one program path. 

\begin{figure*}
\vspace{2mm}
    \centering
    \begin{minipage}{.3\textwidth}
        \centering
        \includegraphics[width=\textwidth]{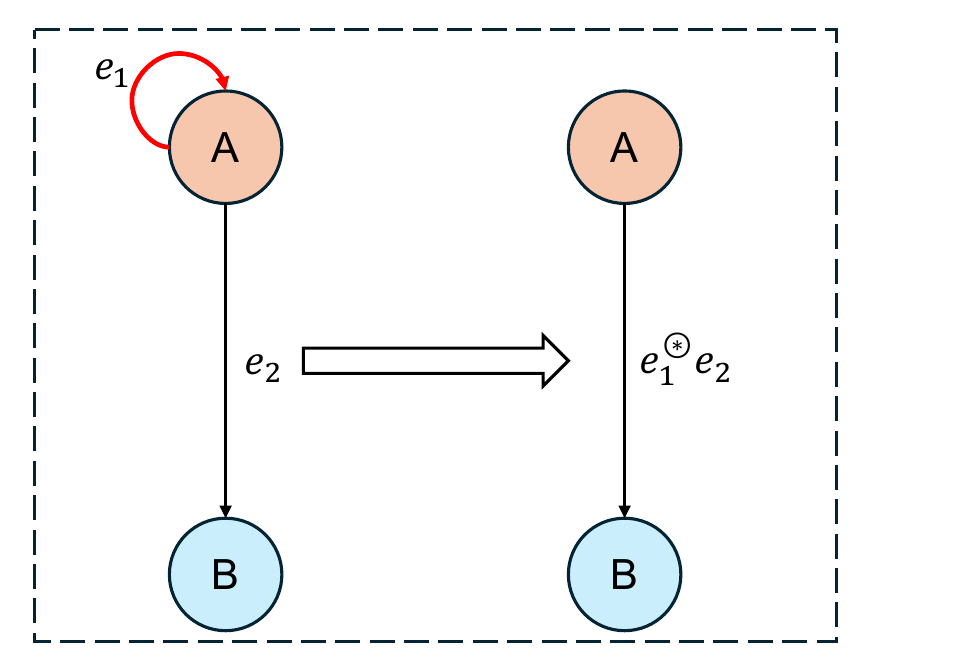}

        {\scriptsize Basic Structure 1}
    \end{minipage}
    \hspace{0.2\textwidth}
    \begin{minipage}{.3\textwidth}
        \centering
        \includegraphics[width=\textwidth]{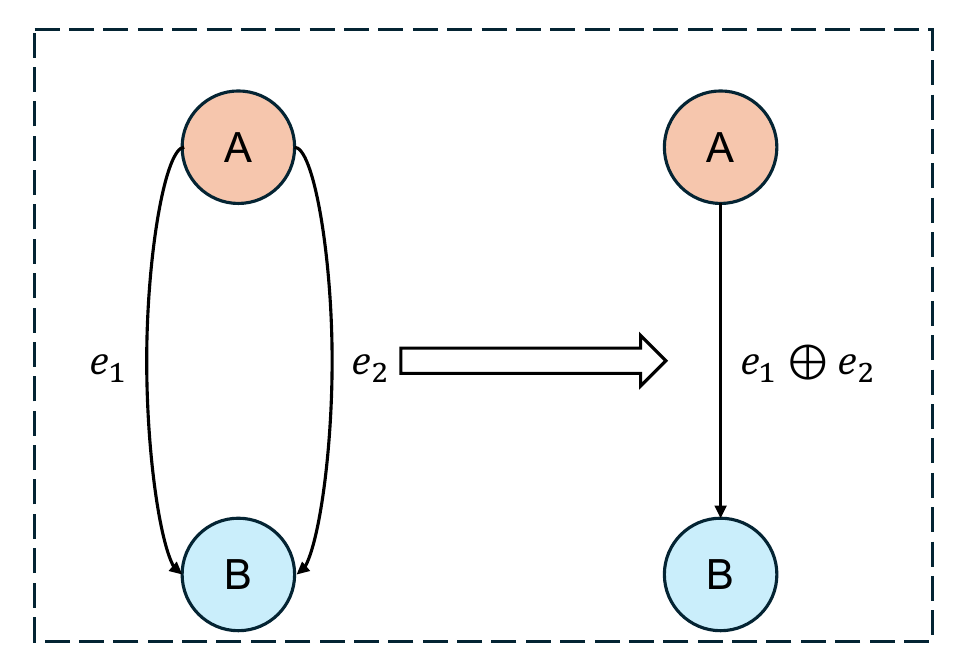}

        {\scriptsize Basic Structure 2}
    \end{minipage}
    \caption{The two basic structures used for hierarchical computation of the path expression.}
    \label{fig:basic_struct}
\end{figure*}

Next, the algorithm applies structural transformations to the program, based on the two pattern-matching-and-replacement rules shown in Fig.~\ref{fig:basic_struct}.  
The first one, called \emph{Basic Structure 1}, removes loops.  The second one, called \emph{Basic Structure 2}, removes branches. 

To efficiently update path expressions, we propose to represent path expression using a \emph{segment tree}, which is a generic data structure for storing information about intervals or segments.  A segment tree can efficiently decide which of the stored segments contains an element . It also maintains a relatively balanced tree, which supports efficient incremental changes.  
%
%
%

Algorithm~\ref{alg:subp-to-tree} shows the baseline (non-incremental) procedure for building the path expression tree, also called the APA-Tree in the remainder of this paper.  The algorithm has a time complexity of $O(n)$, where $n=|\rho|$ is the size of the path expression.  

\begin{algorithm}
\caption{Building the tree $T$ for representing the path expression $\rho$.}
\label{alg:subp-to-tree}
\begin{algorithmic}
{\footnotesize
\Procedure{BuildTree}{ path expression $\rho$ }
\State $T \leftarrow \emptyset$
\If{$|\rho|=1$}
    \State $T.value \gets \rho$    \textcolor{mygreen}{//create a leaf node for an edge $e\in E$ in the control flow graph in the program}
\Else
  \State $mid \gets \frac{|\rho|}{2}$
  \State $T.LeftChild \gets$ \Call{BuildTree}{$\rho_{[1, mid]}$}
  \State $T.RightChild \gets $ \Call{BuildTree}{$\rho_{[mid+1, |\rho|]}$}
  \State $T.Value \gets \text{merge}(T.LeftChild, T.RightChild)$  \textcolor{mygreen}{//creating an internal node for the $\otimes,\oplus$ or $\circledast$ operator}
\EndIf
\State \Return $T$
\EndProcedure
}
\end{algorithmic}
\end{algorithm}

\subsection{The Incremental Algorithm}

We support two classes of program changes: \emph{changes within a subprogram} and \emph{changes in the basic structures} used to combine subprograms, as shown in Fig.~\ref{fig:basic_struct}.
Since a subprogram is a linear sequence of edges, changes within a subprogram are
\begin{itemize}
\item $add(e',e)$ represents adding edge $e$ right after the existing edge $e'$;
\item $delete(e)$ represents deleting edge $e$; and 
\item $update(e)$ represents changing the semantics of edge $e$, e.g., by changing the data flow associated with $e$ but without changing the control flow.
\end{itemize}
Among these three changes, $update(e)$ does not change the path expression at all. Therefore, our incremental algorithm only needs to handle $add(e',e)$ and $delete(e)$.

\myparagraph{Adding  Edges}
Our algorithm takes two steps.  The first step is finding the existing edge $e'$, and the second step is inserting the new edge $e$ after $e'$.  A special case is $e'=\epsilon$ (empty string), when $e$ is inserted at the beginning of the subprogram.
To make the algorithm efficient, we must satisfy the following properties: (1) every internal node is the summary of all leaf nodes of its subtree, and (2)  we must minimize the modification to keep the tree relatively balanced.

Toward this end, we implement $add(e',e)$ by splitting the existing leaf node of $e'$.  
Consider  adding edge $e_5$ in Fig.~\ref{fig:ex-tree2} as an example. In this case, the existing leaf node for $e_4$ is split into two leaf nodes, one for $e_4$ and the other for $e_5$.  At the same time, a new $\otimes$ node is added to represent $e_4\otimes e_5$.
After that, we must traverse the path from the new leaf node $e_5$ to root, and mark all nodes on this path as \emph{modified}.  The reason is because their corresponding program facts may no longer be valid.  In Section~\ref{sec:method_interpret}, we will present our algorithm for incrementally computing the new program facts. 

\myparagraph{Deleting Edges}
Since deleting an edge may drastically change the tree structure, we propose to do so lazily.  That is, $delete(e)$  marks the leaf node $e$ as \emph{deleted}, without actually deleting it from the tree.  Marking $e$ as \emph{deleted} allows our algorithm for interpreting the path expression to ignore this leaf node.  
After that, we must also traverse this leaf-to-root path and mark all nodes on the path as \emph{modified}, since their corresponding program facts may no longer be valid.

The path expression tree (or APA-Tree) is expected to be balanced with a height of $O(\log n)$.  If changes are random, the tree height is expected to remain $O(\log n)$ due to balanced properties of the segment tree.  However, in real-world applications, changes are almost never random, which means that after many rounds of edge addition and deletion, the tree may no longer be balanced.  Thus, we propose to re-balance it when certain conditions are met. \revised{Noticed that there can always be pathological programs with deeply embedded loops, and deeply embedded branches. However, such software code would not be common, and they would be too hard for developers to understand or debug. Thus, we do not optimize for such extreme cases.}

\myparagraph{Re-balancing the Tree}
We maintain a \emph{weight-balanced} tree, which is a binary tree that associates each node with a weight of its subtree.  In other words, each node has the following fields:  node type, node value, left child, right child, and the weight of the subtree. The weight of a node $x$  is defined as the number of leaf nodes inside its subtree.  By definition, the weight satisfies the following property: $x.weight = x.left.weight + x.right.weight$.  
If a node $x$ satisfies $\min(x.left.weight, x.right.weight) \geq \alpha \cdot x.weight$, we say its subtree is \textit{$\alpha$-weight-balanced}, where $0 < \alpha \leq \frac{1}{2}$. 
The height $h$ of a $\alpha$-weighted-balanced tree  satisfies $h \leq \log_{\frac{1}{1-\alpha}} n = O(\log n)$.

Whenever the subtree of a node $x$ is no longer $\alpha$-weight-balanced, we re-balance it.  This is not a trivial task.  Although \emph{rotation} is the most commonly-used  re-balancing strategy in binary search trees, it is not a good strategy for our application.   The reason is because rotation may drastically change the shape of the tree even for small changes to the path expression. 
Thus, we focus on localizing the subtree that needs to be rebalanced. Once we find it, we simply rebuild the tree $T$ from path expression $\rho$ using Algorithm~\autoref{alg:subp-to-tree}. 
Below are the two conditions that trigger re-balancing. 
\begin{itemize}
\item 
After adding an edge, we traverse the leaf-to-root path and find any node $x$ satisfying $\min(x.left.weight, x.right.weight) < \alpha \cdot x.weight$, and rebuild its subtree. 
\item 
After deleting an edge, we traverse the leaf-to-root path and find any node  $x$ satisfying  $x.mark \ge (1-\alpha) \cdot x.weight$, and rebuild its subtree. Here $x.mark$ is the number of marked leaf nodes inside $x$'s subtree.
\end{itemize}
%

\noindent
\revised{In general, every operation that affects the tree structure may invoke the rebuild procedure.}

\myparagraph{Handling Changes in Basic Structures}
Changes in basic structures can also be classified into adding and deleting edges in the basic structures as shown in Fig.~\ref{fig:basic_struct}, where the edge may be the self-loop on the left-hand side, or a branch on the right-hand side. 
Adding an edge in a basic structure leads to a local modification of the path expression tree.  Specifically, in \emph{Basic Structure 1},  adding the edge $e_1$ will change the path expression from $e_2$ to $e_1^*e_2$, meant to be interpreted as $(e_1)^\circledast \otimes e_2$.
In \emph{Basic Structure 2},  adding the edge $e_1$ will change the path expression from $e_2$ to $(e_1+e_2)$, meant to be interpreted as $(e_1\oplus e_2)$.

Deleting an edge in a basic structure is implemented in the same way as deleting an edge within a subprogram.  That is, we mark the tree-node associated with the edge as \emph{deleted}, without actually removing it from the tree. 
However, the semantic function associated with the affected tree-nodes must be changed. Our key idea is to leverage the property of $\mathtt{1}$ and $\mathtt{0}$ in the semantic algebra.  Recall that $\mathtt{1}$ and $\mathtt{0}$ are the maximal and minimal elements in the universe of program facts. We define the following rules for deleting the edge $e$:
\begin{itemize}
    \item $e' \otimes e \rightarrow e \otimes \mathtt{1}$, meaning that for the sequence operator,  we treat $e$ as an unconditional transfer of control, which does not affect the data flow; 
    \item $e' \oplus e \rightarrow e \oplus \mathtt{0}$, meaning that for the choice operator, we treat $e$ as no transfer of control; and 
    \item ${e}^{\circledast} \rightarrow {\mathtt{1}}^{\circledast}$, meaning that for the iteration operator, we treat $e$ as an empty loop. 
\end{itemize}

\section{Incrementally Interpreting Path Expression}
\label{sec:method_interpret}

In this section, we present our method for incrementally interpreting path expression to compute the program facts of interest.

%
In classic APA methods, this problem is equivalent to the expression parsing problem, for which
the most common algorithm relies on two stacks: an operator stack $S_1$ and an operand stack $S_2$. By scanning the regular expression and operating on stacks $S_1$ and $S_2$, one can compute the final program facts.  However, this method is inherently non-incremental; for every newly updated path expression, the program facts are computed from scratch.

\myparagraph{The Modified Nodes in Tree $T$}
Incrementally computing path expression as described in the previous section will modify some leaf-to-root paths in the path expression tree $T$. According to our algorithm, nodes on these modified paths are marked as \emph{modified}.  For each node $x$ in the tree $T$, if the $x.\mathit{modified}$ flag is set, it means the program facts associated with the node must be recomputed.  In contrast, for all other nodes whose \emph{modified} flags are not set, the program facts associated with them are still valid and thus do not need to be recomputed.

\myparagraph{The Universe of Program Facts}
Depending on the actual type of analysis performed by APA, the universe of program facts may vary significantly. Recall that if the goal is to compute the use of possibly-uninitialized variables, the universe of program facts may be defined as $D=(DI,PU)$ as shown in Section~\ref{sec:background}.  

%
Regardless of how $D$ is defined, there is a semantic function for each edge $e\in E$ of the control flow graph, denoted $\mathcal{D}\llbracket  \rrbracket: E\rightarrow D$. Since each $e\in E$ corresponds to a leaf node of the tree $T$, $\mathcal{D}\llbracket e \rrbracket$ returns the set of program facts for the leaf node.
Then, for each internal node of the tree $T$, labeled $\otimes$, $\oplus$ or $\circledast$, there is a semantic function for computing the program facts, by merging the program facts of the child nodes.

\myparagraph{The Incremental Algorithm}
Algorithm~\ref{alg:interpret-exp} shows our procedure for incrementally interpreting the path expression, by recomputing only program facts associated with the modified nodes of the tree $T$.
As shown in the algorithm, if $T.\mathit{modified}$ is false, the current node and its subtree are skipped since its program facts are still valid. We assume that $T.\mathit{fact}$ stores the program facts for the subtree $T$.

\begin{algorithm}[h]
\caption{Incrementally computing program facts for \emph{modified} nodes in path expression tree $T$.}
\label{alg:interpret-exp}
\begin{algorithmic}
{\footnotesize
\Procedure{InterpretTree}{ path expression tree $T$}
\If{$T.\mathit{modified}$} 

\If{$T.\mathit{type} = \mathit{leaf}$}
    \State $T.\mathit{fact} \gets \textsc{computeProgramFacts}(T)$   
              \textcolor{mygreen}{//for each edge $e\in E$ in the control flow graph}
\Else
    \State \Call{InterpretTree}{$T.\mathit{LeftChild}$}
    \State \Call{InterpretTree}{$T.\mathit{RightChild}$}
    \State $T.\mathit{fact} \gets \textsc{mergeProgramFacts}(T.\mathit{LeftChild}.\mathit{fact}, T.\mathit{RightChild}.\mathit{fact})$
              \textcolor{mygreen}{//for $\otimes$, $\oplus$ or $\circledast$ operator}
\EndIf

\EndIf
\State \Return
\EndProcedure
}
\end{algorithmic}
\end{algorithm}

If $T.\mathit{modified}$ is true, the current node and its subtree are processed recursively.  There are two cases. 
The first case is when $T$ is a leaf node.  In this case, we compute program facts for $T.\mathit{fact}$  using the semantic function defined for each edge $e\in E$ of the control flow graph.
The second case is when $T$ is an internal node.  In this case, we compute program facts for $T.\mathit{fact}$ using the semantic functions defined for the $\otimes$, $\oplus$ and $\circledast$ operators.

\section{Mathematical Properties of Incremental APA}
\label{sec:discussion}

In this section, we show that when the semantic algebra satisfies certain conditions, our method for incremental APA has some nice mathematical properties.  We also present the semantic algebras for two other applications: reaching definition analysis and constant-time analysis.

\subsection{Uniqueness of the Analysis Result}

While our method for efficiently conducting incremental APA is applicable to any kind of properly-defined semantic algebras, in general, the analysis result is not unique since it may be affected by the order in which results of subexpressions are combined.   This is somewhat inconvenient in theory and may also become significant in practice.
However, if the semantic algebra is a  Kleene algebra~\cite{kozen1990kleene}, this issue is avoided because Kleene algebra guarantees that the analysis result of APA is unique.  

Furthermore, a large number of practically-important program analysis problems can be implemented using Kleene algebra; they include all three analyses used by our experimental evaluation.  For these analyses, our incremental APA guarantees to return the same result as the baseline (non-incremental) APA while being orders-of-magnitude faster.

\myparagraph{Kleene Algebra}
Kleene algebra is an algebraic system $ \langle A, \otimes, \oplus, \circledast, \mathtt{1}, \mathtt{0} \rangle$ defined as follows.  Given the natural order $\leq$ such that $a \leq b$ iff $a \oplus b = b$, we say that $A$ is a Kleene algebra if the following properties are satisfied for all $a, b, c \in A$:

\begin{itemize}
    \item \emph{{Associativity}}: $a \oplus (b \oplus c) = (a \oplus b) \oplus c$ and $a\otimes (b \otimes c)= (a \otimes b) \otimes c$.
    \item \emph{{Distributivity}}:  $a \otimes (b \oplus c) = (a \otimes b) \oplus (a \otimes c)$ and $(b \oplus c) \otimes a=(b \otimes a) \oplus (c \otimes a)$.
    \item \emph{{Identity}}: $a \oplus \mathtt{0} = a$ and $\mathtt{1} \otimes a = a \otimes \mathtt{1} = a$.
    \item \emph{{Commutativity}}: $a\oplus b = b \oplus a$.
    \item \emph{{Idempotence}}: $a \oplus a = a$.
    \item \emph{{Annihilation}}: $a \otimes \mathtt{0} = \mathtt{0} \otimes a = \mathtt{0}$.
    \item \emph{{Unfolding}}: $\mathtt{1} \oplus a \otimes (a^{\circledast}) = \mathtt{1} \oplus (a^{\circledast})\otimes a = a^{\circledast}$.
    \item \emph{{Induction}}: $a \otimes b \leq b \implies a^{\circledast} \otimes b \leq b$ and $b \otimes a \leq b \implies b \otimes a^{\circledast} \leq b$.
\end{itemize}
%
%
%
The reason why Kleene algebra guarantees that the analysis result is unique is because of the associative law, which makes APA compositional: it allows the program to be divided into components, through the $\otimes$ operator within a subprogram or the $\oplus$ operator outside of a subprogram; furthermore, the order for combining components does not affect the result.

\myparagraph{The Associative Law}
To see how important the associative law is, consider removing it from Kleene algebra and adding the following properties, thus getting a non-associative semi-ring: 
\begin{itemize}
    \item \emph{{Commutative monoid for $\oplus$}}: $a \oplus b = b \oplus a$, $(a \oplus b) \oplus c = a \oplus (b \oplus c)$ and $a \oplus 0 = a$.
    \item \emph{{Magma with unit element for $\otimes$}}: $(a \otimes b) \otimes c \neq a \otimes (b \otimes c)$ and $a \times 1 = a$.
%
%
\end{itemize}
Since this new algebraic system no longer guarantees a commutative monoid for $\otimes$, the corresponding APA must interpret path expression in a certain order.  For example, to compute the use of possibly-uninitialized variables, APA must interpret the left subexpression before the right subexpression.  In this sense, it behaves more like iterative program analysis.

\myparagraph{Pre-Kleene Algebra}
If an algebraic system only satisfies  the first six laws of Kleene algebra, it is called an idempotent semi-ring.  If the unfolding and induction laws of Kleene algebra are replaced by the following weaker properties, the resulting system is called a pre-Kleene algebra: 
\begin{itemize}
    \item \emph{{Reflexivity}}: $1 \leq a^{\circledast}$.
    \item \emph{{Extensivity}}: $a \le a^{\circledast}$.
    \item \emph{{Transitivity}}: $a^{\circledast}\otimes a^{\circledast} = a^{\circledast}$.
    \item \emph{{Monotonicity}}: $a \le b \implies a^{\circledast} \le b^{\circledast}$.
    \item \emph{{Unrolling}}: $\forall n \in \mathbb{N}, (a^{n})^{\circledast} \leq a^{\circledast}$.
\end{itemize}
With pre-Kleene algebra, different path expressions may lead to different analysis results. In a prior work, Cyphert et al.~\cite{cyphert2019refinement} proposed to refine path expression to improve its quality.  Their method is complementary to our method in that, one may use their method to build the path expression tree $T$ and then start using our method to update and interpret $T$ incrementally.

\subsection{Star-free Kleene Algebra}

While studying the mathematical properties of Kleene algebra and its variants, we realize that many practical data-flow analysis problems do not really need the Kleene star (or iteration) operator.  For these analysis problems, the semantic functions for $\otimes$, $\oplus$ and $\circledast$ operators satisfy the following relation:  
$a^\circledast = 1\oplus (a \otimes a)$ for all path expression $a \in A$. 
It means that the impact of iterating through a loop 0 to $+\infty$ times is equivalent to (and thus may be captured by) iterating through the loop exactly twice (denoted $a\otimes a$). 
\begin{itemize}
\item If the loop is skipped all together, the impact is equivalent to $1$; and 
\item after the loop iterates twice ($a\otimes a$), the resulting program facts stop changing, thus making the impact of $a\otimes a$ equivalent to iterating arbitrarily many times. 
\end{itemize}

\myparagraph{Running Example: Computing Reaching Definitions}
A large number of practically-important data-flow analyses can be implemented using semantic algebras that satisfy this property.   Below, we illustrate this property using reaching-definition analysis.  Following Kincaid et al.~\cite{kincaid2021algebraic}, we define the semantic algebra as follows: 
\begin{itemize}
    \item $D=(G,K)$ is the universe of program facts, where $G =2^{Def}$ captures the subsets of generated definitions, and $K= 2^{Def}$ captures the subsets of killed definitions.   In other words, $G$ and $K$ correspond to the gen/kill sets in classic data-flow analysis. 
    \item $\mathcal{D} \lBrack e:x:=t \rBrack \triangleq (\{e\},\{e'|e' \text{ defines }x\})$ is the semantic function for each assignment. 
    \item $\mathcal{D}\lBrack e_1 \times e_2\rBrack := (G_1, K_1) \otimes (G_2, K_2) \triangleq ((G_1\setminus K_2)\cup G_2,(K_1\setminus G_2)\cup K_2)$. 
    \item $\mathcal{D}\lBrack e_1 + e_2\rBrack :=(G_1, K_1) \oplus (G_2, K_2) \triangleq (G_1 \cup G_2, K_1 \cap K_2)$. 
    \item $\mathcal{D} \lBrack e^{*}\rBrack := (G_1, K_1)^{\circledast} = (G, \emptyset)$. 
\end{itemize}
Based on these definitions, we prove $a^\circledast = 1\oplus (a \otimes a)$ holds for all path expression $e_a$ as follows: 
\begin{align}
    \mathcal{D} \lBrack \mathtt{1} + e_a \times e_a \rBrack &= (\emptyset , \emptyset ) \oplus (G, K) \otimes (G, K) \nonumber \\
    &= (\emptyset , \emptyset ) \oplus (G, K) \nonumber\\
    &= (G, \emptyset) = \mathcal{D} \lBrack e_a ^{*} \rBrack \nonumber 
\end{align}

\myparagraph{Star-free Kleene Algebra}
While it is interesting to know that reaching-definition analysis can be implemented without the iteration $\circledast$ operator, a more interesting question (in theory) is how to precisely characterize the class of analyses that satisfy this property.  Toward this end, we take the definition of Kleene algebra and then replace its unfolding law and induction law with the following two star-free laws:
%
\begin{itemize}
    \item \emph{{Star-free Unfolding}}: $\mathtt{1}\oplus a \otimes a = \mathtt{1} \oplus a \oplus a \otimes a \otimes a$.
    \item \emph{{Star-free Induction}}: $a \otimes b \le b \implies b \oplus a \otimes a \otimes b \leq b$ and $b \otimes a \leq b \implies b \oplus b \otimes a \otimes a \leq b$.
\end{itemize}%
%
We name the resulting algebraic system Star-free Kleene Algebra. Program analyses that can be implemented using Star-free Kleene Algebra are guaranteed to be efficient, since the impact of a loop can be computed by iterating through the loop at most twice.

\subsection{Constant-Time Analysis}

We now present another analysis that we have used to experimentally evaluate our method for incremental APA.  The goal is to detect differences in a program's execution time that are also dependent on a secret input. 
At a high level, it can be viewed as a combination of taint analysis and execution time analysis.  The goal of taint analysis is to compute, for a given secret input, the set of program variables whose values may be affected by the secret input.  The goal of execution time analysis is to compute, for all program paths from node $s$ to node $t$ in the control flow graph, the lower and upper bounds of the time taken to execute these paths. 

Since taint analysis is widely known, in the remainder of this section, we focus on the execution time analysis by presenting the semantic algebra for computing upper/lower bounds of a program's execution time. 

\begin{figure}
\vspace{2mm}
\centering
\begin{minipage}{0.36\linewidth}
\centering
\begin{lstlisting}[language=C,firstnumber=1]
   int const_time(int secret) {
    int loop = 0 ,sum = 0, cond;
    int mask= secret - 1;
    if (mask) {
        sum = 0;
    } else cond = 1;
    if (cond > 0) {
        while (loop < 3) {
            sum += 2;
            loop++;
        }
    } else sum--;
    return 0;
 }
\end{lstlisting}
\end{minipage}
\begin{minipage}{0.6\linewidth}
    \centering
    \includegraphics[width=0.9\linewidth]{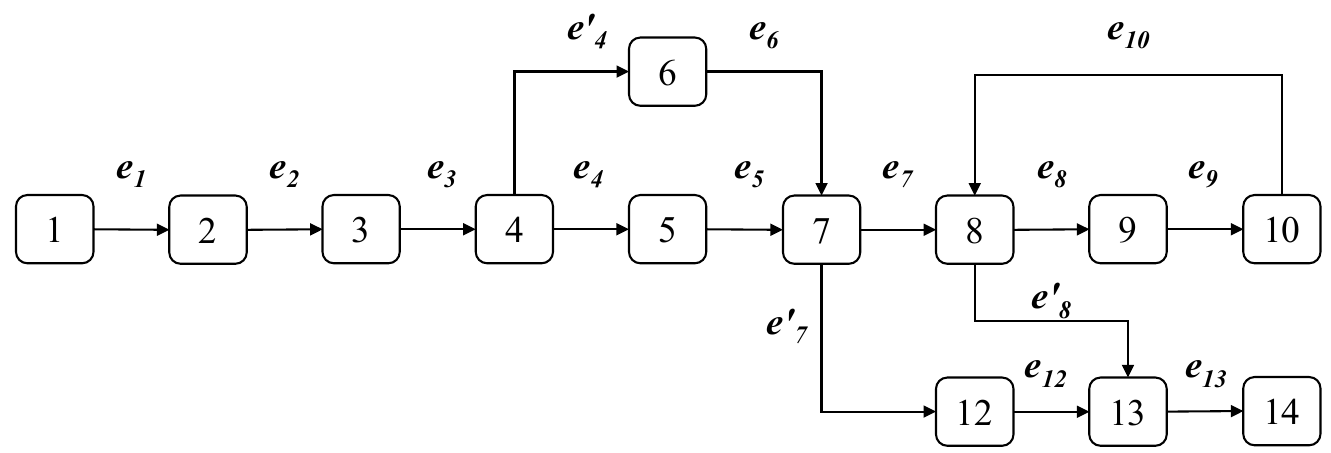}
\vspace{2ex}
\end{minipage}
\caption{An example program for applying APA to compute secret-dependent variation in execution time.}
\label{fig:ex-program3}
\end{figure}

\myparagraph{The Motivating Example}
Fig.~\ref{fig:ex-program3} shows an example program 
\revised{%
with loops and branches; it is a modification of the code snippet from Libgcrypt~\cite{Libgcrypt}, a real cryptographic software program used in prior work~\cite{ZhouDZ24} for evaluating constant-time programming techniques. 
}%
In this program, the secret input affects \texttt{mask} through data-dependency, which in turn affects \texttt{cond} through control-dependency.  Since \texttt{cond} affects which branch in Lines~7-12 is executed, and the execution time of these two branches differ, there is a timing side-channel.  In other words, by observing the execution time, an adversary may gain some information about the secret input.

For ease of understanding, we employ a simple time cost model, by first defining a programming language used to write the program and then assuming that each basic block takes a unit of time to execute. This is without loss of generality, since the time cost model may be replaced by a more accurate one without modifying the algorithm. 
Fig.~\ref{fig:DSL} shows the programming language. 

\begin{figure}
    \centering
{\footnotesize
    \[
\begin{array}{rll}
    \texttt{Arithmetic Expression}~ t &::=& \texttt{constant}~|~t_1+t_2~|~t_1-t_2~|~t_1*t_2 \\
    \texttt{Boolean Expression}~ b &::=&  \texttt{true}~|~\texttt{false}~|~\neg b~|~b_1\wedge~b_2~|~b_1~\vee~b_2 \\
    \texttt{Statement}~ Stmt &::=& x := t~|~\texttt{if}~b~\texttt{then}~Stmt_1~\texttt{else}~Stmt_2~|~\texttt{while}~b~\texttt{do}~Stmt\\
\end{array}
\]
}
    \caption{The definition of a simple programming language used to write the program to be analyzed.}
    \label{fig:DSL}
\end{figure}

\myparagraph{The Semantic Algebra}
The universe of program facts is defined as $D=([LB,UB],C)$, where $LB\in \mathbb{N}$ and $UB\in\mathbb{N}$ represent lower and upper bounds of the execution time, i.e., the  minimum and maximum numbers of basic blocks along program paths of interest, and  $C \in 2^{Var}$ is the set of variables that may control whether the program paths can be executed. 
Regarding the execution time interval $[LB,UB]$:  when $LB=UB$, all program paths have the same execution time, indicating that there is no timing side-channel leakage.  When $LB\neq UB$, however, there is a timing side-channel if the difference in execution time is controlled by the secret input.

The semantic function $\mathcal{D}$, which maps a path expression to a set of program facts in $D=([LB,UB],C)$, is defined as follows\revised{, where $e_b$ represents the expression for the branch condition}:

\[
\begin{array}{rll|rrll}
     \mathcal{D} \lBrack e \rBrack &:=& ([1, 1], \emptyset) &
     \mathcal{D} \lBrack e_b\rBrack &:=& ([0,0], \{x|e_b~\text{uses }x\})\\
     \mathcal{D}\lBrack e_1e_2\rBrack &:=& \mathcal{D}\lBrack e_1\rBrack \otimes \mathcal{D}\lBrack e_2\rBrack \ \ \ \   & \ \ \ \  
     \mathcal{D}\lBrack e_1 + e_2\rBrack &:=& \mathcal{D}\lBrack e_1\rBrack \oplus \mathcal{D}\lBrack e_2\rBrack\\
     \mathcal{D} \lBrack e^{*} \rBrack & := & \mathcal{D} \lBrack e\rBrack^{\circledast} & & &
\end{array}
\]


\vspace{1ex}\noindent
For $L = L_1 \otimes L_2$, the semantic function is defined as follows: 
Let $D_{L_1} = ([LB_1, UB_1], C_1)$ and $D_{L_2} = ([LB_2, UB_2], C_2)$, we have 

\vspace{1ex}
$D_L = ([LB_1+LB_2, UB_1+UB_2], C_1\cup C_2)$. 

\vspace{1ex}\noindent
This is because sequential composition ($\otimes$) will not generate differences in execution time; instead, the total time is the summation of the individual time. 
Furthermore, since all of the variables in $C_1$ and $C_2$ may affect whether we can execute the program path, the two sets are united together.  
%

\vspace{1ex}\noindent
For $L = L_1 \oplus L_2$, the semantic function is defined
%
%
\revised{
by assuming that $T_{L}$ is the precomputed set of secret-tainted variables; we omit the description of taint analysis since it is well understood. 
}%
Let $L_1 = ([LB_1, UB_1], C_1)$  and  $L_2 = ([LB_2, UB_2], C_2)$, we have

\vspace{1ex}
\( D_L = 
    \begin{cases}
      ([\min(LB_1, LB_2), \max(UB_1, UB_2)], C_1\cup C_2) & \text{if } (C_1 \cup C_2) \cap T_L \neq \emptyset \\
       ([LB_1, UB_1], \emptyset) & \text{otherwise}
    \end{cases} \)

\vspace{1ex}\noindent%
%
%
%
%
That is, if the branch statement combining $L_1$ and $L_2$ \revised{(including the branch condition as well as statements inside the then-branch and else-branch)} is controlled by a secret-tainted variable,
it may take different paths based on the secret value, potentially introducing side-channel leaks. In this case, the time interval accounts for both branch executions.
Otherwise, the time difference will not lead to side-channel leakage; thus, we ignore the time difference by arbitrarily taking the time from one branch ($L_1$ in this case). 

\vspace{1ex}\noindent
For $L = L^{\circledast}$, the semantic function is defined as follows:

\vspace{1ex}
\( D_L= 
    \begin{cases}
      ([0, \infty],\emptyset) & \text{if } (C_L) \cap T_{L} \neq \emptyset \\
       ([0, 0], \emptyset) & \text{otherwise}
    \end{cases} \)

\vspace{1ex}\noindent
The situation is similar to $\oplus$: when the loop condition is not controlled by a secret-tainted variable, the time difference will not lead to side-channel leakage; thus, we ignore the time difference by setting the time interval to $[0,0]$;  otherwise, we set the time interval to $[0,\infty]$ to indicate the maximal difference in execution time.

\myparagraph{Applying to the Motivating Example}
Consider the example program in Fig.~\ref{fig:ex-program3}.  
\begin{itemize}
    \item For the $\otimes$ operator, assuming that we have $\mathcal{D}\lBrack e_1e_2e_3(e_4e_5+e_4'e_6) \rBrack = ([4, 4], \{mask\})$ and $\mathcal{D} \lBrack (e_7(e_8e_9e_{10})^*e_8'+e_7'e_{12})e_{13} \rBrack = ([2, \infty], \{cond, loop\})$, based on the semantic function, we get the final result 
    $([6, \infty]$, $\{mask, cond, loop\})$, indicating that there is timing side-channel leakage in this program fragment.
    \item For the $\oplus$ operator, assuming that we have $\mathcal{D}\lBrack e_7(e_8e_9e_{10})^*e_8'\rBrack=([1, \infty], \{cond, loop\})$ and $\mathcal{D}\lBrack e_7'e_{12} \rBrack = ([1, 1], \{cond\})$, since $cond$ is a secret-tainted variable, we maximize the time difference to get the final result  $([1, \infty], \{cond, loop\})$.
    \item For the $\circledast$ operator, assuming that we already have $\mathcal{D} \lBrack e_8e_9e_{10} \rBrack = ([2, 2], \{loop\})$ for the loop body,  thus we get the final result $\mathcal{D} \lBrack e_8e_9e_{10} \rBrack^{\circledast}=([0, \infty], \emptyset)$.
\end{itemize}
At this moment, it is worth noting that the algebra is not a star-free Kleene algebra.  If all the conditional variables used in branching conditions are tainted, the Kleene unfolding rule would still hold here. However, as this above analysis focuses on counting basic blocks, which is significantly different from classic data-flow analysis defined using the gen/kill sets, it is not a star-free analysis. 

\section{Experiments}
\label{sec:experiment}

We have implemented the proposed method in a software tool that conducts algebraic program analysis for Java programs.  The tool builds upon the work of Conrado et al.~\cite{conrado2023exploiting}: it takes the Java bytecode as input and returns the analysis result as output.  Internally, it leverages Soot to construct the control flow graph.  Our implementation of the incremental APA algorithm, consisting of updating the path expression and interpreting the path expression,  is written in 7,640 lines of C++ code.  
To facilitate the experimental comparison, we implemented the baseline APA in the same software tool where we implemented our new method.  In addition, we implemented the method of Conrado et al.~\cite{conrado2023exploiting} and the classic method of Tarjan~\cite{tarjan1981fast}; both of these existing methods were partially implemented as part of the work by Conrado et al.~\cite{conrado2023exploiting}, but their implementation only had the (first) step of computing path expression.  We added the (second) step of interpreting path expression for these two existing methods.  Thus, we are able to have  a fair comparison of all four methods: our incremental APA, baseline APA, the method of Conrado et al.~\cite{conrado2023exploiting}, and the classic method of Tarjan~\cite{tarjan1981fast}.

We implemented APA by defining the semantic algebras and semantic functions to support three types of analyses, consists of two elementary analyses and one compound analysis.  The two elementary analyses are detecting the use of possibly-uninitialized variables and computing reaching definitions. The compound analysis is designed to check whether a program has timing side-channel leakage.   We call it a compound analysis because it consists of a taint analysis and a time analysis: the taint analysis checks if branching conditions are secret-dependent, while the time analysis checks if all branches have the same execution time. 
%

\subsection{Benchmark Statistics}

The benchmark programs used in our experimental evaluation come from the DaCapo~\cite{DaCapo:paper} benchmark suite, consisting of the compiled Java bytecode of 13 real-world applications.  These are open-source applications implementing a diverse set of functionalities. 
Table~\ref{tbl:benchmark} shows the statistics of these benchmark programs. 
Columns~1-3 show the name, the number of lines of Java code, and the number of variables of the program.
Columns~4-5 show the number of nodes and the number of edges of the corresponding control flow graph (CFG). 
Columns~6-7 show the number of definitions  and the number of uses of program variables.  Here, a definition means a write to a variable by a program statement, and a use means a read from a variable.

\begin{table}
\centering
\caption{Statistics of the benchmark programs.
}
\label{tbl:benchmark}
\scalebox{0.78}{
\begin{tabular}{|l|r|r|r|r|r|r|}
\hline
Program   & \# LoC     & \# Variables   & \# CFG Nodes  & \# CFG Edges  & \# Definitions     & \# Uses    \\\hline\hline
hsqldb    &  23,362 & 12,352       & 23,362     & 24,739     & 15,946           & 47,690   \\\hline
avrora    &  87,173 & 47,449       & 87,173     & 85,681     & 60,816           & 172,410  \\\hline
xalan     &  22,314 & 11,780       & 22,314     & 23,595     & 15,268           & 45,555   \\\hline
pmd       & 191,672 & 100,688      & 191,672    & 198,773    & 127,223          & 385,537  \\\hline
fop       & 220,726 & 119,419      & 220,726    & 221,381    & 163,890          & 459,354  \\\hline
luindex   &  41,372 & 20,815       & 41,372     & 42,907     & 28,692           & 80,503   \\\hline
bloat     &  81,557 & 46,140       & 81,557     & 84,408     & 55,239           & 161,960  \\\hline
jython    & 107,115 & 56,830       & 107,115    & 107,450    & 70,551           & 207,161  \\\hline
lusearch  & 47,123  & 23,403       & 47,123     & 48,848     & 33,030           & 90,702   \\\hline
eclipse   & 73,869  & 39,826       & 73,869     & 76,309     & 47,334           & 145,280  \\\hline
antlr     & 43,563  & 23,847       & 43,563     & 45,383     & 29,121           & 88,013   \\\hline
sunflow   & 70,769  & 38,869       & 70,769     & 72,376     & 55,364           & 159,305  \\\hline
chart     & 149,926 & 79,085       & 149,926    & 152,246    & 104,081          & 310,694  \\\hline
\end{tabular}
}
\end{table}

\revised{
Changes made to benchmark programs are generated automatically by following established practice~\cite{szabo2016inca,szabo2021incremental}. Since we use Soot to transform Java bytecode of each program to a CFG before conducting APA, program instruction-level changes correspond to CFG edge insertions, deletions, and edge content modifications.
Note that unlike a standard CFG where nodes are labeled with basic blocks, in our CFG, basic blocks are moved from nodes to their outgoing edges as illustrated in Fig.~\ref{fig:ex-program}. Thus, we randomly generate program changes until the number of affected basic blocks reaches 2\%, 4\%, 6\%, \dots, 20\%. These percentages are consistent with our stated goal of conducting APA incrementally in response to \emph{small and frequent changes} of the program.
}

\revised{More specifically, we generate changes by randomly selecting one basic block at a time, and then performing the following operations: (i) delete the basic block, which may affect both control and data flows; (ii) add assignment statements in the basic block, which will be chosen from existing or new program variables with different possibilities and may affect the data flow; and (iii) add a new basic block behind the chosen basic block with randomly assigned data flow facts, which may affect both control and data flows. 
}

\subsection{Experimental Setup}

We conducted all of our experiments on a computer with Intel Core i7-8700 CPU and  32GB memory, running the Ubuntu 22.04 operating system. 
Our experiments were designed to answer the following research questions (RQs):
\begin{enumerate}[label=RQ \arabic*., leftmargin=*]
\item
Is our method for conducting APA incrementally significantly more efficient than the baseline APA and existing methods of Conrado et al.~\cite{conrado2023exploiting} and Tarjan~\cite{tarjan1981fast}?
\item
How does each of the two technical innovations (incrementally computing path expression and incrementally interpreting path expression) contribute to the overall performance improvement?
\item
How does the size of program change affect the performance improvement of our method? 
\end{enumerate}

%

\subsection{Results for RQ 1}

To answer RQ 1, we experimentally compared the performance of all four methods (incremental APA, baseline APA, Conrado et al.~\cite{conrado2023exploiting} and Tarjan~\cite{tarjan1981fast}) on all benchmark programs.

Table~\ref{tbl:result} shows the results, divided into three subtables for the three types of analyses.
Column~1 shows the program name. 
Columns~2-4 show the result of incremental APA, including the size of the path expression tree $T$, the size of $F$ for storing program facts, and the total analysis time in milliseconds (ms).
Columns~5-6 show the time of baseline APA and, in comparison, the percentage of time taken by incremental APA; the percentage equals the time of incremental APA divided by the time of baseline APA. 
Columns~7-10 show the time of classic method of Tarjan~\cite{tarjan1981fast}, the time of recent method of Conrado et al.~\cite{conrado2023exploiting}, and the percentages of time taken by incremental APA.

\begin{table}
\centering
\caption{Comparing our method with the baseline (non-incremental) and two existing APA methods~\cite{tarjan1981fast,conrado2023exploiting} for solving three types of program analysis problems.
}
\label{tbl:result}

{\footnotesize \bf
\vspace{-1ex}
Reaching Definitions 
\vspace{1ex}}

\scalebox{0.78}{
\begin{tabular}{|l|c|c|r|r|c|r|c|r|c|}
\hline
Program   & \multicolumn{3}{c|}{Incremental APA } & \multicolumn{2}{c|}{Baseline APA} & \multicolumn{2}{c|}{Tarjan~\cite{tarjan1981fast}}   & \multicolumn{2}{c|}{Conrado et al.~\cite{conrado2023exploiting}} \\\cline{2-10}
          & expression size & property size & time (ms)  & time (ms)    & \%  & time (ms) & \% &  time (ms) & \%  \\\hline\hline
antlr & 19,324 & 24,161 & 6.921 & 319.969 & 2.163 & 328.199 & 2.109 & 30,290.090 & 0.023 \\\hline
luindex & 10,388 & 11,819 & 1.159 & 84.743 & 1.367 & 86.720 & 1.336 & 62,226.830 & 0.002 \\\hline
avrora & 2,174 & 2,458 & 0.222 & 7.734 & 2.868 & 8.177 & 2.712 & 330.622 & 0.067 \\\hline
jython & 50,321 & 60,175 & 14.177 & 2,957.460 & 0.479 & 2,972.265 & 0.477 & 263,940.250 & 0.005 \\\hline
fop & 4,469 & 5,004 & 0.945 & 43.714 & 2.161 & 44.666 & 2.115 & 1,978.554 & 0.048 \\\hline
lusearch & 13,683 & 15,697 & 2.039 & 119.962 & 1.699 & 123.460 & 1.651 & 72,641.133 & 0.003 \\\hline
pmd & 34,405 & 42,173 & 9.211 & 2,974.718 & 0.310 & 2,988.893 & 0.308 & 393,436.950 & 0.002 \\\hline
xalan & 1,055 & 1,149 & 0.274 & 10.392 & 2.636 & 10.683 & 2.564 & 280.962 & 0.097 \\\hline
chart & 56,462 & 66,055 & 1.578 & 162.644 & 0.970 & 179.290 & 0.880 & 25,993.126 & 0.006 \\\hline
hsqldb & 897 & 984 & 0.138 & 8.750 & 1.575 & 9.155 & 1.506 & 48.702 & 0.283 \\\hline
bloat & 58,552 & 69,893 & 3.601 & 226.068 & 1.593 & 242.902 & 1.483 & 25,164.876 & 0.014 \\\hline
eclipse & 17,859 & 20,638 & 0.824 & 42.021 & 1.962 & 46.672 & 1.766 & 6,599.907 & 0.012 \\\hline
sunflow & 49,240 & 56,270 & 2.688 & 63.922 & 4.205 & 80.829 & 3.326 & 17,111.078 & 0.016

\\\hline\hline
{\bf Total} & 318,829 & 376,476 & {\bf 43.777}  & {\bf 7,022.096}  & {\bf 0.623}  &7,121.912 & {\bf 0.615} & 900,043.082 & {\bf 0.005} \\\hline
\end{tabular}
}

{\footnotesize \bf
\vspace{2ex}
Use of Possibly-uninitialized Variables
\vspace{1ex}
}

\scalebox{0.78}{
\begin{tabular}{|l|c|c|r|r|c|r|c|r|c|}
\hline
Program   & \multicolumn{3}{c|}{Incremental APA } & \multicolumn{2}{c|}{Baseline APA} & \multicolumn{2}{c|}{Tarjan~\cite{tarjan1981fast}}   & \multicolumn{2}{c|}{Conrado et al.~\cite{conrado2023exploiting}} \\\cline{2-10}
          & expression size & property size & time (ms)  & time (ms)    & \%  & time (ms) & \% &  time (ms) & \%  \\\hline\hline
antlr & 19,324 & 24,161 & 4.745 & 288.278 & 1.646 & 207.883 & 2.283 & 30,123.315 & 0.016 \\\hline
luindex & 10,388 & 11,819 & 0.794 & 82.782 & 0.960 & 64.519 & 1.231 & 57,625.330 & 0.001 \\\hline
avrora & 2,174 & 2,458 & 0.208 & 6.952 & 2.987 & 5.361 & 3.874 & 365.905 & 0.057 \\\hline
jython & 50,321 & 60,175 & 10.228 & 2,907.694 & 0.352 & 2,733.111 & 0.374 & 305,885.858 & 0.003 \\\hline
fop & 4,469 & 5,004 & 0.562 & 31.848 & 1.764 & 12.087 & 4.647 & 1,716.860 & 0.033 \\\hline
lusearch & 13,683 & 15,697 & 1.320 & 110.936 & 1.190 & 84.267 & 1.566 & 73,156.096 & 0.002 \\\hline
pmd & 34,405 & 42,173 & 4.338 & 2,680.677 & 0.162 & 2,637.441 & 0.164 & 257,063.222 & 0.002 \\\hline
xalan & 1,055 & 1,149 & 0.161 & 14.269 & 1.129 & 2.796 & 5.762 & 286.831 & 0.056 \\\hline
chart & 56,462 & 66,055 & 1.080 & 145.929 & 0.740 & 116.453 & 0.927 & 27,280.602 & 0.004 \\\hline
hsqldb & 897 & 984 & 0.072 & 11.251 & 0.644 & 0.930 & 7.785 & 50.366 & 0.144 \\\hline
bloat & 58,552 & 69,893 & 1.827 & 204.626 & 0.893 & 169.574 & 1.077 & 25,410.865 & 0.007 \\\hline
eclipse & 17,859 & 20,638 & 0.504 & 38.048 & 1.323 & 28.605 & 1.760 & 6,776.800 & 0.007 \\\hline
sunflow & 49,240 & 56,270 & 2.086 & 49.031 & 4.254 & 41.122 & 5.072 & 17,921.532 & 0.012 

\\\hline\hline
{\bf Total} & 318,829 & 376,476 &  {\bf 27.924}  & {\bf 6,572.321}  & {\bf 0.425}  &6,104.149 &  {\bf 0.457} & 803,663.583 & {\bf 0.003} \\\hline
\end{tabular}
}

{\footnotesize \bf
\vspace{2ex}
Constant-time Execution 
\vspace{1ex}
}

\scalebox{0.78}{
\begin{tabular}{|l|c|c|r|r|c|r|c|r|c|}
\hline
Program   & \multicolumn{3}{c|}{Incremental APA } & \multicolumn{2}{c|}{Baseline APA} & \multicolumn{2}{c|}{Tarjan~\cite{tarjan1981fast}}   & \multicolumn{2}{c|}{Conrado et al.~\cite{conrado2023exploiting}} \\\cline{2-10}
          & expression size & property size & time (ms)  & time (ms)    & \%  & time (ms) & \% &  time (ms) & \%  \\\hline\hline

antlr & 19,324 & 24,161 & 1.128 & 2,462.752 & 0.046 & 2,470.732 & 0.046 & 30,310.704 & 0.004 \\\hline
luindex & 10,388 & 11,819 & 0.321 & 1,803.192 & 0.018 & 1,805.190 & 0.018 & 57,360.517 & 0.001 \\\hline
avrora & 2,174 & 2,458 & 0.111 & 327.153 & 0.034 & 327.660 & 0.034 & 352.924 & 0.031 \\\hline
jython & 50,321 & 60,175 & 3.089 & 22,223.328 & 0.014 & 22,243.659 & 0.014 & 280,295.457 & 0.001 \\\hline
fop & 4,469 & 5,004 & 0.195 & 50.960 & 0.382 & 52.478 & 0.371 & 2,153.981 & 0.009 \\\hline
lusearch & 13,683 & 15,697 & 0.386 & 491.880 & 0.078 & 497.042 & 0.078 & 84,238.675 & 0.000 \\\hline
pmd & 34,405 & 42,173 & 1.885 & 5,294.618 & 0.036 & 5,308.758 & 0.036 & 286,184.554 & 0.001 \\\hline
xalan & 1,055 & 1,149 & 0.074 & 1,285.112 & 0.006 & 1,285.595 & 0.006 & 288.575 & 0.026 \\\hline
chart & 56,462 & 66,055 & 0.317 & 4,023.013 & 0.008 & 4,039.288 & 0.008 & 27,269.490 & 0.001 \\\hline
hsqldb & 897 & 984 & 0.078 & 1,442.620 & 0.005 & 1,443.114 & 0.005 & 45.678 & 0.170 \\\hline
bloat & 58,552 & 69,893 & 0.373 & 345.759 & 0.108 & 369.157 & 0.101 & 27,002.183 & 0.001 \\\hline
eclipse & 17,859 & 20,638 & 0.193 & 210.871 & 0.092 & 214.518 & 0.090 & 6,265.510 & 0.003 \\\hline
sunflow & 49,240 & 56,270 & 0.252 & 581.690 & 0.043 & 601.567 & 0.042 & 18,015.339 & 0.001 

\\\hline\hline
{\bf Total} & 318,829 & 376,476 & {\bf 8.402}  &{\bf 40,542.948}  & {\bf 0.021}  &40,658.756 & {\bf 0.021} & 819,783.588 & {\bf 0.001}  \\\hline 
\end{tabular}
}
\end{table}

In addition to the individual results for the 13 benchmark programs,  Table~\ref{tbl:result} also shows the aggregated total at the bottom of the three subtables. 
Overall, our method is significantly faster than the other three methods.
For reaching definitions, the total time of incremental APA is 43.777 ms, which is only 0.623\% of the 7,022.096 ms taken by baseline APA.  In other words, the speedup is 160$\times$.
For the use of possibly-uninitialized variables and constant-time analysis, the speedup is 235$\times$ and 4761$\times$, respectively.
The result demonstrates the effectiveness of our proposed techniques for incrementally computing path expression and computing program properties.

The result also shows that our baseline APA is as competitive as the fast APA method of Tarjan; they have similar total analysis time (7,022.096 ms versus 7,121.912  ms). 
In contrast, the time taken by the most recent APA method of Conrado et al.~\cite{conrado2023exploiting} is significantly longer (900,043.082 ms); this should not be surprising because the method was optimized for solving a different problem, i.e., how to answer a large number of queries in nearly constant time for a fixed program.  Toward this end, the  method shares the cost of answering all queries by precomputing a lot of information, which slows down the computation of path expression by more than 100 times.

For these experiments, the size of program change was set to 4\%, although due to the nature of the three different analyses, the 4\% change applied to the programs are different for each analysis. 
Since existing APA methods were not designed to efficiently handle program changes, even for a slightly modified program, these methods would have to recompute the path expression from scratch, whereas our method directly updates the existing path expression. This is the reason why our method takes only a tiny fraction of the time. 
%

\subsection{Results for RQ 2}

To answer RQ 2, we conducted an ablation study by measuring performance improvement contributed by each of the two components of incremental APA: the method for incrementally computing path expression and the method for incrementally interpreting path expression.

Table~\ref{tbl:breakdown} shows the breakdown of analysis time taken by our method and the baseline APA.  
Column~1 shows the program name. 
Columns~2-4 show the time taken by our method to update path expression, the time taken to update program facts, and the total time in milliseconds (ms).  
Columns~5-7 show the corresponding time taken by the baseline APA. 
Columns~8-10 show the percentage of time taken by our method  compared with the baseline APA (time for updating path expression, time for updating program properties, and the total).

\begin{table}
\centering
\caption{The breakdown of analysis time taken by our method and baseline (non-incremental) APA.
}
\label{tbl:breakdown}

{\footnotesize \bf
\vspace{-1ex}
Reaching Definitions 
\vspace{1ex}}

\scalebox{0.78}{
\begin{tabular}{|l|c|c|c|c|c|c|c|c|c|}
\hline
Program & \multicolumn{3}{|c|}{Incremental APA (ms)} &\multicolumn{3}{c|}{Baseline APA (ms)} & \multicolumn{3}{c|}{ Ratio (\%) } \\\cline{2-10}
& tree  & fact  & total   
& tree  & fact  & total   
& tree  & fact  & total   
\\\hline\hline 

antlr & 0.795 & 6.127 & 6.921 & 199.513 & 120.456 & 319.969 & 0.40 & 5.09 & 2.16 \\\hline
luindex & 0.238 & 0.921 & 1.159 & 60.083 & 24.659 & 84.743 & 0.40 & 3.73 & 1.37 \\\hline
avrora & 0.071 & 0.151 & 0.222 & 5.201 & 2.533 & 7.734 & 1.36 & 5.96 & 2.87 \\\hline
jython & 1.908 & 12.269 & 14.177 & 2,711.405 & 246.055 & 2,957.460 & 0.07 & 4.99 & 0.48 \\\hline
fop & 0.136 & 0.809 & 0.945 & 10.837 & 32.877 & 43.714 & 1.25 & 2.46 & 2.16 \\\hline
lusearch & 0.265 & 1.773 & 2.039 & 80.701 & 39.260 & 119.962 & 0.33 & 4.52 & 1.70 \\\hline
pmd & 1.576 & 7.634 & 9.211 & 2,727.554 & 247.165 & 2,974.718 & 0.06 & 3.09 & 0.31 \\\hline
xalan & 0.060 & 0.214 & 0.274 & 2.252 & 8.141 & 10.392 & 2.65 & 2.63 & 2.64 \\\hline
chart & 0.193 & 1.385 & 1.578 & 95.976 & 66.667 & 162.644 & 0.20 & 2.08 & 0.97 \\\hline
hsqldb & 0.032 & 0.106 & 0.138 & 0.599 & 8.151 & 8.750 & 5.34 & 1.30 & 1.58 \\\hline
bloat & 0.330 & 3.271 & 3.601 & 151.456 & 74.612 & 226.068 & 0.22 & 4.38 & 1.59 \\\hline
eclipse & 0.135 & 0.689 & 0.824 & 24.106 & 17.915 & 42.021 & 0.56 & 3.85 & 1.96 \\\hline
sunflow & 0.171 & 2.517 & 2.688 & 23.896 & 40.027 & 63.922 & 0.72 & 6.29 & 4.21 \\\hline\hline

\textbf{Total} & {\bf 5.910} & {\bf 37.866} & 43.777 & {\bf 6,093.579} & {\bf 928.517} &  7,022.096 & {\bf 0.10} & {\bf 4.08} & 0.62   \\\hline
\end{tabular}
}

{\footnotesize \bf
\vspace{2ex}
Use of Possibly-uninitialized Variables
\vspace{1ex}
}

\scalebox{0.78}{
\begin{tabular}{|l|c|c|c|c|c|c|c|c|c|}
\hline
Program & \multicolumn{3}{|c|}{Incremental APA (ms)} &\multicolumn{3}{c|}{Baseline APA (ms)} & \multicolumn{3}{c|}{ Ratio (\%) } \\\cline{2-10}
& tree  & fact  & total   
& tree  & fact  & total   
& tree  & fact  & total   
\\\hline\hline 

antlr & 0.734 & 4.011 & 4.745 & 200.143 & 88.135 & 288.278 & 0.37 & 4.55 & 1.65 \\\hline
luindex & 0.220 & 0.574 & 0.794 & 62.547 & 20.235 & 82.782 & 0.35 & 2.84 & 0.96 \\\hline
avrora & 0.060 & 0.147 & 0.208 & 4.872 & 2.080 & 6.952 & 1.24 & 7.08 & 2.99 \\\hline
jython & 1.716 & 8.511 & 10.228 & 2,718.280 & 189.414 & 2,907.694 & 0.06 & 4.49 & 0.35 \\\hline
fop & 0.105 & 0.457 & 0.562 & 11.099 & 20.750 & 31.848 & 0.95 & 2.20 & 1.76 \\\hline
lusearch & 0.250 & 1.069 & 1.320 & 80.823 & 30.113 & 110.936 & 0.31 & 3.55 & 1.19 \\\hline
pmd & 0.967 & 3.371 & 4.338 & 2,623.012 & 57.665 & 2,680.677 & 0.04 & 5.85 & 0.16 \\\hline
xalan & 0.051 & 0.110 & 0.161 & 2.485 & 11.784 & 14.269 & 2.04 & 0.94 & 1.13 \\\hline
chart & 0.164 & 0.915 & 1.080 & 97.342 & 48.587 & 145.929 & 0.17 & 1.88 & 0.74 \\\hline
hsqldb & 0.027 & 0.045 & 0.072 & 0.583 & 10.668 & 11.251 & 4.62 & 0.43 & 0.64 \\\hline
bloat & 0.246 & 1.580 & 1.827 & 153.182 & 51.443 & 204.626 & 0.16 & 3.07 & 0.89 \\\hline
eclipse & 0.122 & 0.381 & 0.504 & 24.207 & 13.841 & 38.048 & 0.51 & 2.75 & 1.32 \\\hline
sunflow & 0.162 & 1.923 & 2.086 & 24.052 & 24.979 & 49.031 & 0.68 & 7.70 & 4.25 \\\hline\hline

\textbf{Total} & {\bf 4.827} & {\bf 23.097} & 27.924 & {\bf 6,002.627} & {\bf 569.694} & 6,572.321 & {\bf 0.08} & {\bf 4.05} & 0.42 \\\hline
\end{tabular}
}

{\footnotesize \bf
\vspace{2ex}
Constant-time Execution 
\vspace{1ex}
}

\scalebox{0.78}{
\begin{tabular}{|l|c|c|c|c|c|c|c|c|c|}
\hline
Program & \multicolumn{3}{|c|}{Incremental APA (ms)} &\multicolumn{3}{c|}{Baseline APA (ms)} & \multicolumn{3}{c|}{ Ratio (\%) } \\\cline{2-10}
& tree  & fact  & total   
& tree  & fact  & total   
& tree  & fact  & total   
\\\hline\hline 

antlr & 1.057 & 0.070 & 1.128 & 293.041 & 2,169.711 & 2,462.752 & 0.36 & 0.00 & 0.05 \\\hline
luindex & 0.304 & 0.017 & 0.321 & 88.838 & 1,714.354 & 1,803.192 & 0.34 & 0.00 & 0.02 \\\hline
avrora & 0.104 & 0.007 & 0.111 & 11.480 & 315.673 & 327.153 & 0.91 & 0.00 & 0.03 \\\hline
jython & 2.929 & 0.161 & 3.089 & 3,789.976 & 18,433.352 & 22,223.328 & 0.08 & 0.00 & 0.01 \\\hline
fop & 0.169 & 0.026 & 0.195 & 16.204 & 34.756 & 50.960 & 1.04 & 0.07 & 0.38 \\\hline
lusearch & 0.365 & 0.021 & 0.386 & 129.647 & 362.232 & 491.880 & 0.28 & 0.01 & 0.08 \\\hline
pmd & 1.754 & 0.131 & 1.885 & 3,730.134 & 1,564.484 & 5,294.618 & 0.05 & 0.01 & 0.04 \\\hline
xalan & 0.070 & 0.004 & 0.074 & 3.136 & 1,281.976 & 1,285.112 & 2.24 & 0.00 & 0.01 \\\hline
chart & 0.300 & 0.016 & 0.317 & 161.927 & 3,861.086 & 4,023.013 & 0.19 & 0.00 & 0.01 \\\hline
hsqldb & 0.076 & 0.001 & 0.078 & 0.911 & 1,441.709 & 1,442.620 & 8.38 & 0.00 & 0.01 \\\hline
bloat & 0.349 & 0.025 & 0.373 & 224.864 & 120.896 & 345.759 & 0.16 & 0.02 & 0.11 \\\hline
eclipse & 0.180 & 0.013 & 0.193 & 60.170 & 150.701 & 210.871 & 0.30 & 0.01 & 0.09 \\\hline
sunflow & 0.233 & 0.018 & 0.252 & 74.617 & 507.074 & 581.690 & 0.31 & 0.00 & 0.04 \\\hline\hline

\textbf{Total} & {\bf 7.891} & {\bf 0.510} & 8.402 & {\bf 8,584.943} & {\bf 31,958.004} & 40,542.948 & {\bf 0.09} & {\bf 0.00} & 0.02 \\\hline
\end{tabular}
}
\end{table}

Overall, both components in our method contributed to the significant reduction in analysis time. 
For example, during incremental APA for reaching definitions, the total analysis time of 43.777 ms is divided into 5.910 ms for updating path expression  and 37.866 ms for updating  program facts.  This is in sharp contrast to the total analysis time of 7,022.096 ms for the baseline APA, divided into 6,093.579 ms for computing path expression from scratch and 928.517 ms for computing program facts from scratch.

For reaching definition, while the overall speedup (7022.096 ms / 43.777 ms) is 160$\times$, the speedup on computing path expression (6093.579 ms / 5.910 ms) is 1031$\times$ and the speedup on computing program properties (928.517 ms / 37.866 ms) is 24$\times$. 

For the use of possibly-uninitialized variables, while the overall speedup (8701.310 ms / 48.799 ms) is 178$\times$, the speedup on computing path expression (6452.497 ms / 29.610 ms) is 217$\times$ and the speedup on computing program properties (2248.813 ms / 19.189 ms) is 117$\times$. 

For constant-time analysis, while the overall speedup (40542.948 ms / 8.402 ms) is 4825$\times$, the speedup on computing path expression (8584.943 ms / 7.891 ms) is 1087$\times$ and the speedup on computing program properties (31958.004 ms / 0.510 ms) is 62662$\times$.

%
For constant-time execution, the reason why incrementally updating program properties (facts) has a much larger speedup is because its abstract domain has more dimensions than the other two analyses.  This leads to a much bigger time cost for the baseline APA to recompute for the unchanged parts of the program. 
%

\subsection{Results for RQ 3}

\begin{figure}
\centering
\begin{minipage}{\linewidth} 
\begin{minipage}{0.495\linewidth}
\centering
    \includegraphics[width=1\linewidth]{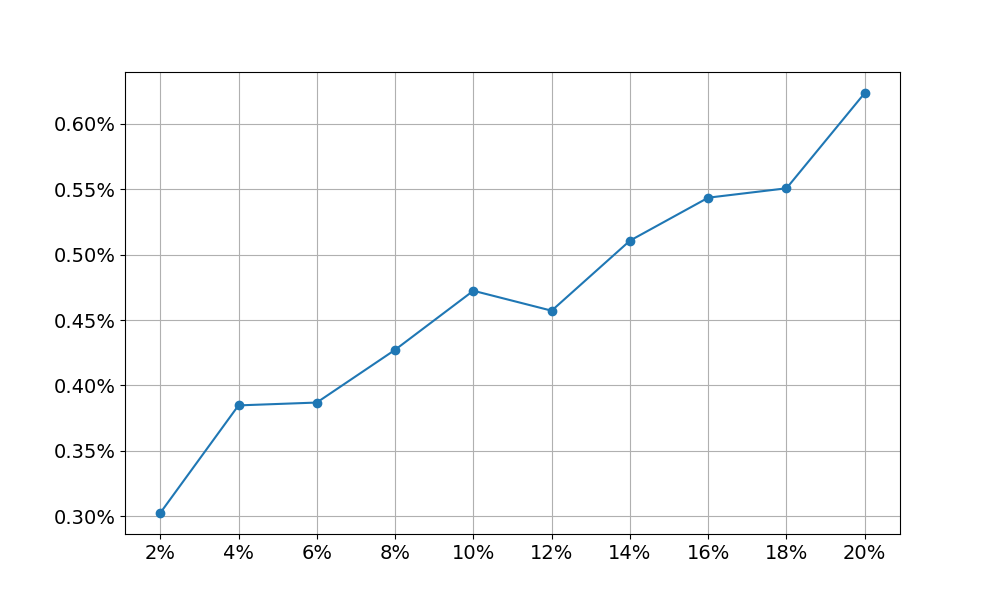}
{\scriptsize (a) time for updating tree}
\end{minipage}
\begin{minipage}{0.495\linewidth}
\centering
    \includegraphics[width=1\linewidth]{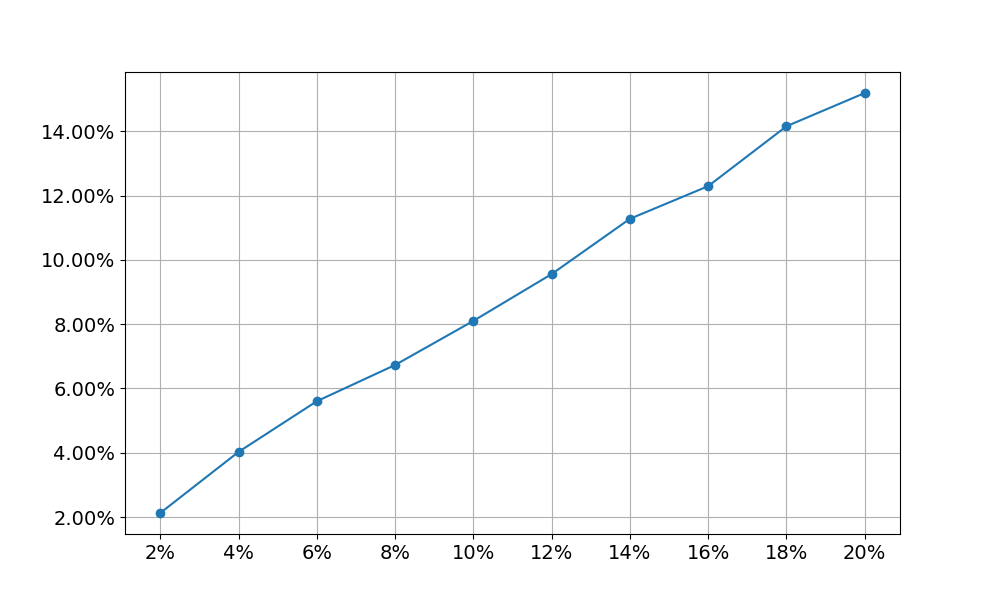}
{\scriptsize (b) time for updating facts}
\end{minipage}
\caption{Reaching definitions for the program \textcolor{black}{\texttt{luindex}}: analysis time of incremental APA  where $x$-axis is the size of program change (0--20\%) and $y$-axis is the reduced time (percentage w.r.t.\ baseline APA time).}
\label{fig:time-change-reaching-def}
\end{minipage}
\end{figure}

The experimental results presented in the previous subsections were obtained using 4\% of program change during incremental APA.  To answer RQ 3, we evaluate how the percentage of program change affects the performance of incremental APA.  Toward this end, we varied the percentage of program change within the range of 2\% -- 20\% and measured the analysis time.  
Fig.~\ref{fig:time-change-reaching-def} shows the results of this experiment for computing reaching definitions conducted on the benchmark program named \texttt{luindex}.

In this figure, the $x$-axis represents the percentage of program changes in the range of 0--20\%, and the $y$-axis represents the reduced time (i.e., the percentage of time w.r.t. baseline APA) for updating path expression and updating program facts, respectively. 
For example, when the size of program change increases from 2\% to 20\%, the time for updating path expression increases from  0.30\% (of the time taken for computing path expression from scratch) to 0.65\%. 
At the same time, the time for updating program facts increases from 2\% (of the baseline time) to 15\%.

\begin{figure}
\centering
\begin{minipage}{\linewidth} 
\begin{minipage}{0.495\linewidth}
\centering
   \includegraphics[width=1\linewidth]{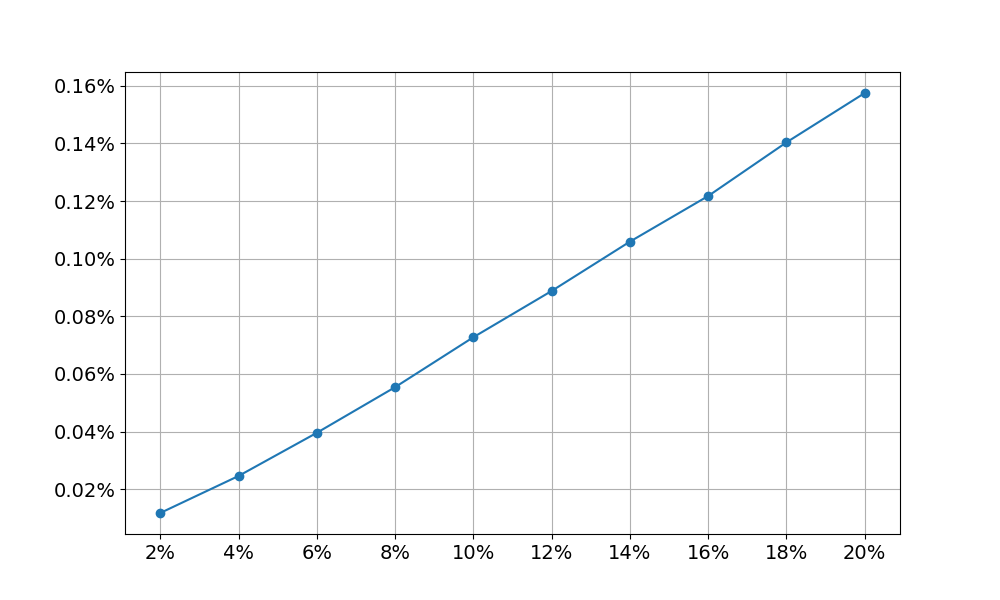}

{\scriptsize (a) updated tree nodes}
\end{minipage}
\begin{minipage}{0.495\linewidth}
\centering
   \includegraphics[width=1\linewidth]{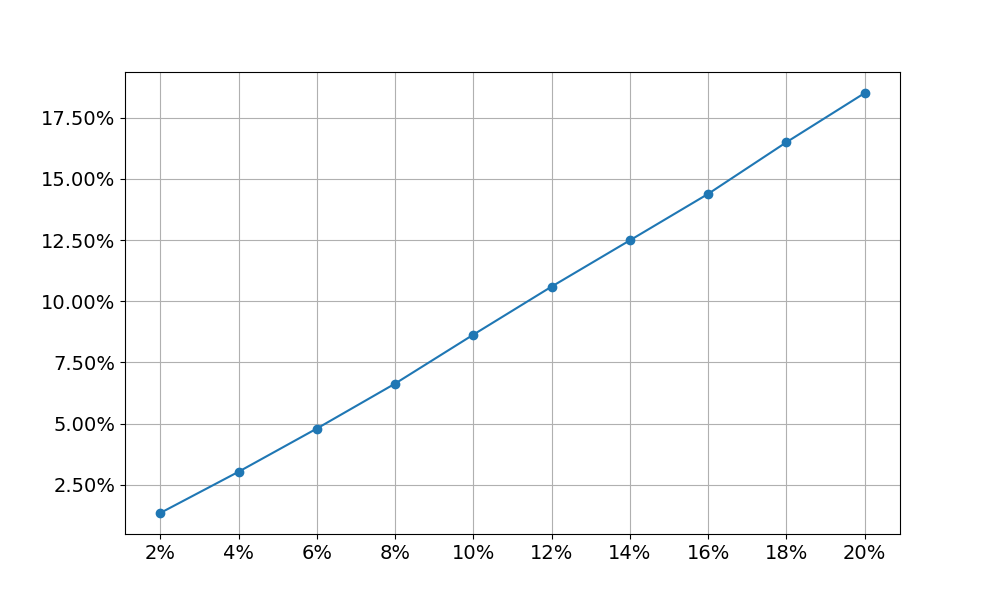}

{\scriptsize (b) updated facts}
\end{minipage}
\caption{Reaching definitions for the program \textcolor{black}{\texttt{luindex}}: updated nodes in path expression and facts by incremental APA,  where $x$-axis is the size of program change (0--20\%) and $y$-axis is the updated tree nodes and program facts (percentage w.r.t.\ baseline APA).}
\label{fig:node-change-reaching-def}
\end{minipage}
\end{figure}

To understand the reason why the time changes in the way shown by Fig.~\ref{fig:time-change-reaching-def}, we also measured the percentage of updated nodes in the path expression tree $T$ and the percentage of nodes whose associated facts are also updated.  Recall that, for the running example in Fig.~\ref{fig:ex-tree2}, these updated nodes and facts are highlighted in red color; our hypothesis is that, when the program change is small, the updated tree nodes and facts are also small, and they are the reasons why our method for incremental APA can have a significant reduction in analysis time. 
The results of this experiment, shown in Fig.~\ref{fig:node-change-reaching-def}, confirm our hypothesis. 
When the size of program change increases from 2\% to 20\%, the percentage of updated nodes in the path expression increases from 0.02\% to 0.16\%, and the percentage of updated program facts increases from 1.25\% to 18\%.

\revised{%
The reason why we keep the program changes below 20\% is because the goal is to conduct APA incrementally in response to small and frequent changes. Above 20\%, the changes are no longer small.  
That said, out of curiosity, we have conducted experiments with the program changes increased to 100\%. At 100\%, the curves in Fig.~\ref{fig:time-change-reaching-def} would go slightly above 100\%, indicating that incremental APA is slower than baseline APA because removing all nodes from the existing APA-tree and then rebuilding the tree from scratch) take time. The curves in Fig.~\ref{fig:node-change-reaching-def} reach exactly 100\%, indicating that all nodes and facts are recomputed. 
}

\section{Related Work}
\label{sec:related}

As mentioned earlier, our method is the first method for conducting APA incrementally in response to program changes.  While there is a large body of work on APA in the literature, to the best of our knowledge, none of the existing methods leverages the intermediate results computed for a previous version of the program to speed up APA for the current program. 
For example, while we experimentally compared with two existing methods (in addition to baseline APA), these existing methods were not designed to solve the same problem targeted by our method.

In particular, the method of Conrado et al.~\cite{conrado2023exploiting} was designed to quickly answer a large number of queries on a fixed version of the program, by amortizing the computational cost only during the first step of APA (which is computing the path expression); they did not mention nor implement the second step (which is interpreting the path expression). 
%
Since we reused their tool for experimental comparison, we had to implement the second step for their method, to facilitate a fair comparison with our method and other APA methods. 

While we are not aware of any existing work on incremental APA, there is a large body of work on classic (non-incremental) APA.  These classic techniques can be traced all the way back to Tarjan's fast algorithm for computing path expression~\cite{tarjan1981fast} and his unified framework~\cite{tarjan1981unified} for solving path problems. 
Reps et al.~\cite{reps1995precise} were the first to compute path expression in polynomial time, essentially by leveraging Tarjan's algorithm.
For more information about classic and recent techniques on APA, please refer to the tutorial paper by Kincaid et al.~\cite{kincaid2021algebraic}.

Improving the efficiency of APA is only one of the related research directions. Another research direction is improving the quality of path expression computed from the control flow graph.  Recall that path expression guarantees to capture all feasible program paths of interest, but in order to be efficient, it may also capture some infeasible program paths.  Generally speaking, the fewer infeasible program paths, the better, since infeasible program paths lead to less accurate analysis result.  
Cyphert et al.~\cite{cyphert2019refinement} developed a technique for refining  a given path expression to improve its quality: by reducing the number of infeasible program paths captured by the path expression, they were able to improve the accuracy of APA.

Besides classic data-flow analyses, APA has been used in a wide range of applications including but not limited to  invariant generation~\cite{KincaidCBR18}, termination analysis~\cite{ZhuKFY21}, invariant generation~\cite{kincaid2018numerical}, predicate abstraction~\cite{RepsTP16}, and more recently, the analysis of probabilistic programs~\cite{WangHR18,wang2023newtonian}.
Among these applications, a particularly interesting line of work is to exploit the inherent compositionality of APA~\cite{farzan2013algebraic,farzan2015compositional}. Being compositional means that the result of analyzing a program can be computed from results of analyzing the individual components in isolation.  This is important because compositionality allows APA to scale to large programs and to be easily parallelized.

Beyond APA, the problem of incrementally updating analysis results has been studied in the context of iterative program analysis, e.g., by Ryder~\cite{ryder1983incremental}, Pollock and Soffa~\cite{PollockS89}, and Arzt and Bodden~\cite{arzt2014reviser}.
The declarative program analysis framework~\cite{WhaleyL04,LamWLMACU05} that has become  popular in recent years~\cite{WangSRW21,WangSW19,PaulsenSPW19,SungLEW18,SungKW17} can also be viewed as a form of iterative program analysis, for which the fixed-point computation is performed either by BDDs~\cite{NaikAW06} or a Datalog solver~\cite{AntoniadisTS17}. Incremental algorithms have been proposed for these declarative program analysis techniques~\cite{szabo2016inca,szabo2021incremental,ZhaoSRS21}.

\section{Conclusion}
\label{sec:conclusion}

We have presented a method for incrementally conducting algebraic program analysis in response to changes of the program under analysis. 
Compared to the baseline APA, our method consists of two new components.  The first component is designed to represent the path expression as a tree and to efficiently update the tree in response to program changes. The second component is designed to efficiently update program facts of interest in response to changes of the path expression. Overall, the goal is to reduce the analysis time by leveraging intermediate results that are already computed for the program before changes are made. 
Our experimental evaluation on 13 real-world Java applications from the DaCapo benchmark suite shows that our method is hundreds to thousands of times faster than the baseline APA and two other existing methods.

\section*{Acknowledgments}

This research was supported in part by the U.S.\ National Science Foundation (NSF) under grant CCF-2220345. 
We thank the anonymous reviewers for their constructive feedback.

\bibliographystyle{ACM-Reference-Format}
\bibliography{src/main}

\end{document}